\documentclass[usenatbib]{mn2e}
\usepackage{amsmath,graphics,amssymb}
\usepackage{times}

\allowdisplaybreaks

\newfont{\eufont}{eufm10}

\newcommand{\dmdt}{\ensuremath{\dot{M}}}
\newcommand{\ms}{\ensuremath{\text{M}_{\odot}}}
\newcommand{\msr}{\ensuremath{\ms\,\text{yr}^{-1}}}
\newcommand{\kms}{\ensuremath{\mathrm{km}\,\mathrm{s}^{-1}}}
\newcommand\fx{\ensuremath{f_\text{X}}}

\newcommand{\vr}{{v_r}}

\newcommand{\zav}[1]{\left(#1\right)}
\newcommand{\hzav}[1]{\left[#1\right]}

\newcommand{\de}{\mathrm{d}}
\newcommand{\Teff}{\mbox{$T_\mathrm{eff}$}}

\newcommand\NR{N\!R}
\newcommand\IR{I\!R}
\newcommand\urel{\ensuremath{u_\text{rel}}}
\newcommand\rfir{\ensuremath{R_{f\!i\!r}}}

\defcitealias{pulchuch}{P06}
\defcitealias{martclump}{M05}
\defcitealias{rosat}{BSC}
\defcitealias{oskal}{OC}
\DeclareMathAlphabet{\mathsc}{OT1}{cmr}{m}{sc}
\def\testbx{bx}%
\DeclareRobustCommand{\ion}[2]{%
\relax\ifmmode
\ifx\testbx\f@series
{\mathbf{#1\,\mathsc{#2}}}\else
{\mathrm{#1\,\mathsc{#2}}}\fi
\else{#1\,{\scshape{#2}}}%
\fi}
\title[NLTE models of line-driven stellar winds III.]
      {NLTE models of line-driven stellar winds\\%
      III. Influence of X-ray radiation on wind structure of O stars}

\author[J.  Krti\v{c}ka and J. Kub\'at]{Ji\v r\'\i\  Krti\v{c}ka$^{1}$\thanks{E-mail: krticka@physics.muni.cz (JKr); kubat@sunstel.asu.cas.cz (JKu)
	} and Ji\v r\'\i\  Kub\'at$^{2}$%
\footnotemark[1]\\
	$^{1}$\'Ustav teoretick\'e fyziky a astrofyziky,
        Masarykova univerzita,
	CZ-611 37 Brno, Czech Republic\\
	$^{2}$Astronomick\'y \'ustav
	AV \v{C}R, Fri\v{c}ova 298,
	CZ-251 65 Ond\v{r}ejov, Czech Republic}

\begin{document}

\date{Received}

\maketitle

\begin{abstract}
We study the influence of X-rays on the wind structure of selected O stars. For
this purpose we use our non-local thermodynamic equilibrium (NLTE) wind code with
inclusion of additional artificial source of X-rays, assumed to originate in the
wind shocks.

We show that the influence of shock X-ray emission on wind mass-loss rate is
relatively small. Wind terminal velocity may be slightly influenced by the
presence of strong X-ray sources, especially for stars cooler than
$T_\text{eff}\lesssim35\,000\,$K.

We discuss the origin of the $L_\text{X}/L\sim10^{-7}$ relation. For stars with
thick wind this relation can be explained assuming that the cooling time depends
on wind density. Stars with optically thin winds exhibiting the "weak wind
problem" display enhanced X-ray emission which may be connected with large shock
cooling length.
We propose that this effect can explain the "weak wind problem".

Inclusion of X-rays leads to a better agreement of the model ionization
structure with observations. However, we do not found any significant influence
of X-rays on \ion{P}{v} ionization fraction implying that the presence of X-rays
cannot explain the \ion{P}{v} problem.

We study the implications of modified ionization equilibrium due to shock
emission on the line transfer in the X-ray region. We conclude that the X-ray
line profiles of helium-like ions may be affected by the line absorption within
the cool wind.
\end{abstract}


%________________________________________________________________

\section{Introduction}

The evolution of hot stars seems to be strongly influenced by the loss of a
substantial part of their mass via winds. Consequently, a proper knowledge of
amount of material expelled from the stellar surface per unit of time (mass loss
rate) is necessary for the stellar evolution calculations.

There exist some serious problems with mass-loss rate predictions for hot stars.
The situation is not satisfactory even in the O star domain, where several
different theoretical models are available (e.g., Vink et al. \citeyear{vikola},
Pauldrach et al. \citeyear{pahole}, \citealt{pureho}, \citealt[hereafter
Paper~I]{nltei}). A good agreement between these theoretical models and
observations may be just illusory effect of the possible manifestation of local
wind inhomegeneites (clumping) in the observed spectra (e.g., Bouret et
al.~\citeyear{bourak}, Martins et al. \citeyear{martclump}, \citealt{pulchuch}).
The problem is that the spectrum of a clumped wind with small mass-loss rate may
mimic the spectrum of a smooth wind with large mass-loss rate. Consequently, the
mass-loss rates derived from observations with account of clumping may be
significantly lower than those derived without taking into account the influence
of clumping. Lower mass-loss rates are indicated also by weak \ion{P}{v} wind
lines \citep{fuj}, and by the study of X-ray line profiles \citep{ocpor}. If the
mass-loss rates derived from observations are really much lower due to the
influence of clumping on the observed spectra, then there exists a significant
discrepancy between theory and observation \citep{pulamko}.

Moreover, some O stars with low luminosities exhibit the so-called "weak wind
problem" \citep{bourak,martin, martclump, nlteii}. Theoretically predicted
mass-loss rates of these stars are much higher than the mass-loss rates inferred
from observations (even without taking into account the clumping). The origin of
this discrepancy is unclear.

One of the reasons that may cause these mentioned differences between
observation and theory of hot star winds is the neglect of X-ray radiation. The
X-ray radiation of hot stars is likely generated by the wind shocks that exist
in the wind either due to the instability caused by the radiative driving
(Owocki et al.~\citeyear{ocr}, Feldmeier et al.~\citeyear{felpulpal}) or due to
the collisions of wind streams channelled by the magnetic field \citep[ud-Doula
\& Owocki \citeyear{udo}]{bamo}. The X-rays may be generated also due to
decoupling of wind components accompanied by frictional heating \citep{kapusty}.
In the case of binaries the X-rays may also originate in wind collisions
\citep{usaci, igor}.

X-ray radiation influences above all trace ionization states in most of hot O
stars (MacFarlane et al. \citeyear{macown}) and the inclusion of X-ray sources
is necessary especially for the correct prediction of hot star spectra (Pauldrach
et al. \citeyear{pahole}). On the other hand, for cooler stars the wind
ionization balance may be dominated by the influence of X-ray radiation
(MacFarlane et al. \citeyear{macown}). 

Recently \cite{fuj} showed that the mass-loss rate determined from the
\ion{P}{v} lines is much lower than that derived from the H$\alpha$ line or
radio emission. This implies also that hot star wind mass-loss rates determined
from \ion{P}{v} lines are significantly lower than the theoretical ones.
However, this conclusion is sensitive to the ionisation state of phosphorus.
%Kr1: If the actual phosphorus ionization state is significantly different from
%that assumed 
If \ion{P}{v} is not a dominant ion at any effective temperature studied
by \citet [e.g., due to the presence of X-rays] {fuj}, then the
discrepancy between theory and observation of OB star winds may not be so
significant. Hence, the inclusion of X-ray ionization may also be important from
this perspective.

Thus, we present here a study of the influence of the X-ray radiation on
the wind properties of O stars.

\section{NLTE wind models}

The process of calculation of models used in this paper was described in
Paper~I, thus here we only summarise its basic features and describe
improvements.

\subsection{Basic model description}

Our models assume spherically symmetric stationary stellar wind.

Excitation and ionization state of elements important for the radiative
driving and for the correct calculation of the radiative field is
derived from the statistical equilibrium (NLTE) equations.

The radiative transfer equation is artificially split (as in the Paper I) into
two parts, namely the radiative transfer in continuum and the radiative transfer
in lines. The solution of the radiative transfer in continuum is based on the
Feautrier method in the spherical coordinates (Mihalas \& Hummer
\citeyear{sphermod} or Kub\'at \citeyear{dis}) with inclusion of all free-free
and bound-free transitions of model ions, however neglecting line transitions.
The radiative transfer in lines is solved in the Sobolev approximation (e.g.,
Castor \citeyear{cassob}) neglecting continuum opacity and line overlaps.

The radiative force (calculated in the Sobolev approximation after
(I.25)\footnote{The equations from the Paper~I are denoted as I.$x$, where $x$
is the equation number there.} using data extracted in 2002 from the VALD
database (Piskunov et al. \citeyear{vald1}, Kupka et al. \citeyear{vald2})) and
the radiative cooling/heating term (we use the electron thermal balance method,
Kub\'at et al., \citeyear{kpp}) are calculated using occupation numbers derived
from statistical equilibrium equations. Finally, the continuity equation,
equation of motion, and energy equation are solved and consistent wind velocity,
density, and temperature structure are obtained using iteration procedure. For
our calculations we use \citet{asgres} solar abundance determinations.

The lowest wavelength considered in the models is $7.5$\,\AA\ and the largest
X-ray wavelength is defined as $100$\AA.

\subsection{Inclusion of shock X-ray radiation}
\label{inclsok}

The consistent inclusion of X-ray generation into NLTE wind models (using
hydrodynamical simulations that are able to predict the shock properties,
Feldmeier et al.~\citeyear{felpulpal}) is likely beyond the possibilities of
present computers. To make the problem more tractable, and since we are
interested in the effect of already generated X-rays on the wind, not in the
process of X-ray generation itself, we use a simpler approach after Pauldrach et
al. (\citeyear{lepsipasam}), i.e., we include X-rays into our stationary models
in an artificial way using two free parameters introduced below.

We \emph{assume}
that a part of wind material is heated to a very high temperature
$T_\text{X}$ due to the shock. The shock temperature is given by the
Rankine-Hugoniot shock condition
\begin{equation}
\label{tx}
T_\text{X}=\frac{3m_\text{H}}{32k} \hzav{u_\text{X}^2+ \frac{14}{5}a_\text{H}^2
\zav{1-\frac{3}{14}\frac{a_\text{H}^2}{u_\text{X}^2}}},
\end{equation}
where $k$ is the Boltzmann constant, $m_\text{H}$ is the hydrogen mass,
$a_\text{H}$ is the sound speed calculated assuming
a completely ionised hydrogen plasma,
\begin{equation}
a_\text{H}=\sqrt{\frac{10}{3}\frac{kT_\text{H}}{m_\text{H}}},
\end{equation}
where $T_\text{H}$ is the hydrogen temperature taken from stationary wind
models, and $u_\text{X}$ is the upstream shock velocity,
\begin{equation}
\label{urel} u_\text{X}=\urel \vr,
\end{equation}
where $\vr$ is the radial velocity in the stellar rest frame and $\urel$ is a
dimensionless free parameter influencing the hardness of X-rays. Instead of the
linear dependence of $u_\text{X}$ on the wind velocity it would be possible to
employ different assumptions, e.g., constant $u_\text{X}$. However, the observed
X-rays originate at larger radii where the wind velocity approaches the terminal
one, consequently such dependence would not result in a very different results.
Moreover, in the case of constant $u_\text{X}$ it would be necessary to
introduce additional free parameter corresponding to the radius at which X-rays
start to be emitted. Without it we would obtain unrealistically strong X-ray
source at the wind base as the wind density is the largest there.

We add the X-ray emissivity
\begin{equation}
\label{etax}
\eta_\text{X}(\nu)=\fx\zav{\xi\rho}^2\,\Lambda_\nu(T_\text{X})/\zav{4\pi},
\end{equation}
into the emission coefficient, where $\fx$ is the second free parameter 
determining the amount of X-rays (also called the filling factor),
$\Lambda_\nu(T_\text{X})$ is calculated using the Raymond-Smith X-ray
spectral code (Raymond \& Smith \citeyear{rs}, Raymond \citeyear{ray}), and
$\xi$ relates the wind density $\rho$ and the electron number density 
\begin{equation}
\label{xne}
\xi =\frac{1}{m_\text{H}} \frac{1+2y}{1+4y},
\end{equation}
where $y$ is the helium number density relative to the hydrogen one (here we
assume a fully ionised gas).

We used different values of \fx\ in our calculations. The free parameter
$\urel=0.3$ in all presented calculations. This roughly gives the same X-ray
temperature compared to the results of numerical simulations (e.g., Owocki et
al.~\citeyear{ocr}, Feldmeier et al.~\citeyear{felpulpal}, Runacres \&
Owocki~\citeyear{runow})
%Kr1:
and observations \citep[e.g.,][]{rosatvel,milca,sane}.

\subsection{Joint calculation of wind model}

In our previous work we split the calculation of wind model to parts below and
above the critical point, at which the mass-loss rate of our stationary wind
models is determined \citep{cak}. This effectively means that we neglected the
influence of the region above the critical point onto the region below this
point. This was justifiable in our previous models with negligible contribution
of X-ray radiation, because the wind close the critical point is optically thick
in the UV region and optically thin for other wavelengths (except lines). The
splitting of wind solution may not be legitimate in the case where additional
sources of X-rays are present, because the X-rays that predominantly originate
in the wind above the critical point, may penetrate downwards and, consequently,
influence the solution below this point.

Here we calculate the joint wind model below and above the critical point
already during the procedure of mass-loss rate determination. To do so, we first
calculate several global iteration steps (during which we search for the base
density corresponding to the solution which smoothly passes through the critical
point, see Paper~I) for models in which we account only for the wind close to
the stellar surface. As soon as the appropriate base density is roughly known
(with a precision of about 30\%), we add the solution above the critical point
and perform several additional global iteration steps until the base density is
known with sufficiently high precision of $1\%$. This approach enables us to
properly take into account the influence of X-rays on the mass-loss rate.

\subsection{Statistical equilibrium equations}

The statistical equilibrium equation for level $i$ of a given atom (ion) has the
form (Mihalas \citeyear{mihalas}, also (I.1))
\begin{equation}
\label{nlte}
\sum_{j\neq i}N_j P_{ji}-N_i \sum_{j\neq i}P_{ij}=0,
\end{equation}
where $N_i$, $N_j$ are relative occupation numbers of studied levels
($N_i=n_i/n_\text{atom}$, where $n_i$ is the number density of atoms in given
excitation and ionization state $i$ and $n_\text{atom}$ is the total number
density of given atoms; similarly for the level $j$). $P_{ij}$ are rates of all
processes by which an atom can change its state (see e.g., Mihalas
\citeyear{mihalas}). Note that for simplicity we do not explicitly write the
radial dependence of all variables in equation~\eqref{nlte}.

\subsubsection{Modification of atomic data}

\begin{table}
\caption{Atoms and {ions} included in the NLTE calculations. Here `Level'
means either an individual level or a set of levels merged into a superlevel.}
\label{prvky}
\centering
\begin{tabular}{*{3}{l@{\hspace{-0.7mm}}r@{\hspace{7mm}}}l@{\hspace{-0.7mm}}r}
\hline
Ion & Levels & Ion & Levels & Ion & Levels & Ion & Levels\\
\hline
\ion{H}{i}   &  9& \ion{O}{ii}  & 50& \ion{Al}{ii}  & 16& \ion{Ar}{iii} & 25\\ 
\ion{H}{ii}  &  1& \ion{O}{iii} & 29& \ion{Al}{iii} & 14& \ion{Ar}{iv}  & 19\\ 
\ion{He}{i}  & 14& \ion{O}{iv}  & 39& \ion{Al}{iv}  & 14& \ion{Ar}{v}   & 16\\ 
\ion{He}{ii} & 14& \ion{O}{v}   & 14& \ion{Al}{v}   & 16& \ion{Ar}{vi}  & 11\\ 
\ion{He}{iii}&  1& \ion{O}{vi}  & 20& \ion{Al}{vi}  &  1& \ion{Ar}{vii} &  1\\ 
\ion{C}{ii}  & 14& \ion{O}{vii} &  1& \ion{Si}{ii}  & 12& \ion{Ca}{ii}  & 16\\ 
\ion{C}{iii} & 23& \ion{Ne}{ii} & 15& \ion{Si}{iii} & 12& \ion{Ca}{iii} & 14\\ 
\ion{C}{iv}  & 25& \ion{Ne}{iii}& 14& \ion{Si}{iv}  & 13& \ion{Ca}{iv}  & 20\\ 
\ion{C}{v}   & 11& \ion{Ne}{iv} & 12& \ion{Si}{v}   & 15& \ion{Ca}{v}   & 22\\ 
\ion{C}{vi}  &  1& \ion{Ne}{v}  & 17& \ion{Si}{vi}  &  1& \ion{Ca}{vi}  &  1\\ 
\ion{N}{ii}  & 14& \ion{Ne}{vi} & 11& \ion{P }{iii} & 16& \ion{Fe}{iii} & 29\\ 
\ion{N}{iii} & 32& \ion{Ne}{vii}&  1& \ion{P }{iv}  & 17& \ion{Fe}{iv}  & 32\\ 
\ion{N}{iv}  & 23& \ion{Na}{ii} & 13& \ion{P }{v}   & 21& \ion{Fe}{v}   & 30\\ 
\ion{N}{v}   & 13& \ion{Na}{iii}& 14& \ion{P }{vi}  & 14& \ion{Fe}{vi}  & 27\\ 
\ion{N}{vi}  & 15& \ion{Na}{iv} & 18& \ion{P }{vii} &  1& \ion{Fe}{vii} &  1\\ 
\ion{N}{vii} &  1& \ion{Na}{v}  & 16& \ion{S }{ii}  & 14& \ion{Ni}{iii} & 36\\ 
             &   & \ion{Na}{vi} &  1& \ion{S }{iii} & 10& \ion{Ni}{iv}  & 38\\ 
             &   & \ion{Mg}{iii}& 14& \ion{S }{iv}  & 18& \ion{Ni}{v}   & 48\\ 
             &   & \ion{Mg}{iv} & 14& \ion{S }{v}   & 14& \ion{Ni}{vi}  &  1\\ 
             &   & \ion{Mg}{v}  & 13& \ion{S }{vi}  & 16\\
             &   & \ion{Mg}{vi} &  1& \ion{S }{vii} &  1\\                  
\hline
\end{tabular}
\end{table}

Model atoms are based on set of TLUSTY files (Hubeny \citeyear{tlusty}, Hubeny
\& Lanz \citeyear{hublaj}, Hubeny \& Lanz \citeyear{hublad}, Lanz \& Hubeny
\citeyear{lahub}). The original set described in Paper~I was extended to allow
for consistent inclusion of X-rays (see Table~\ref{prvky}). Our model atoms are
based on the data derived from the Opacity Project (Seaton \citeyear{top},
\citeauthor{opc5} \citeyear{opc5}, Luo \& Pradhan \citeyear{top1}, Sawey \&
Berrington \citeyear{savej}, Seaton et al. \citeyear{topt}, Butler et al.
\citeyear{bumez}, Nahar \& Pradhan \citeyear{napra}) and Iron Project (Hummer et
al. \citeyear{zel0}, Bautista \citeyear{zel6}, Nahar \& Pradhan \citeyear{zel2},
Zhang \citeyear{zel1}, Bautista \& Pradhan \citeyear{zel5}, Zhang \& Pradhan
\citeyear{zel4}, Chen \& Pradhan \citeyear{zel3}). For phosphorus we employed
data described by Pauldrach et al. (\citeyear{pahole}).

\subsubsection{Auger ionization}

In addition to the direct ionization (of a valence electron), the Auger
ionization can significantly alter the ionization state of a stellar wind in the
presence of X-rays (see Cassinelli \& Olson \citeyear{casol}, Olson \& Castor
\citeyear{olca}, MacFarlane et al. \citeyear{macown}, Pauldrach et al.
\citeyear{lepsipasam}).

To include the Auger ionization we inserted Auger photoionization terms 
into statistical equilibrium equations
\eqref{nlte},
\begin{equation}
\label{augernlte}
\sum_{j\neq i}N_j P_{ji}-N_i \sum_{j\neq i}P_{ij}
-N_i \sum_{j>i} R_{ij}^{\text{Auger}}
=0,
\end{equation}
where $R_{ij}^{\text{Auger}}$ is the Auger photoionization rate
with $j$ corresponding to the ground level of higher ions.
Introducing the notation $\text{ion}(i)$, which means the ionization
state, to which the level $i$ belongs,
the Auger photoionization rate is given as a product
\begin{equation}
\label{augerclen}
R_{ij}^{\text{Auger}}=a_{\text{ion}(i)\text{ion}(j)} R_{\text{ion}(i)}^{\text{Auger}},
\end{equation}
where $a_{\text{ion}(i)\text{ion}(j)}$ is the Auger yield, i.e., the probability
that $\text{ion}(j)-\text{ion}(i)$ electrons are expelled due to the Auger
ionization of ionic state $\text{ion}(i)$, i.e. that ionization state
$\text{ion}(j)$ is created during the process of Auger ionization, and
$R_{\text{ion}(i)}^{\text{Auger}}$ is the total inner-shell photoionization rate
of $\text{ion}(i)$. The total inner-shell photoionization rate is given as a sum
of partial inner-shell photoionization rates from all closed inner shells. Auger
rates for transitions ending in the ionization states which are not included in
statistical equilibrium equations are assumed to contribute to the closest lower
ionization state considered. The term corresponding to the Auger ionization is
also included in the absorption coefficient in the continuum part of the
radiative transfer equation.

The influence of Auger ionization on the temperature and the photon emission due
to Auger ionization are neglected.

Photoionization cross sections from individual inner-shells were taken from
Verner \& Yakovlev (\citeyear{veryak}, see also Verner et al.
\citeyear{muzustahnouthura}) and Auger yields were taken from Kaastra \& Mewe
(\citeyear{kame}).

\subsubsection{Accelerated lambda iterations}

Iterative solution of the radiative transfer equation together with the
statistical equilibrium equations may in some cases cause numerical problems
with convergence (see Hubeny \citeyear{hubtueb} for a review). To avoid these
problems, we included accelerated lambda iterations into our models. Their
inclusion is based on the method proposed by Rybicki \& Hummer
(\citeyear{rybashumremali}). The linearization of derived equations is based on
Newton-Raphson iterations. Thus, our method resembles approximate Newton-Raphson
method of Hempe \& Sch\"onberg (\citeyear{hemso}). 
The statistical equilibrium equation \eqref{nlte} have for each $i$ the form of
\begin{multline}
\label{poradtrebova}
\sum_{j<i}4\pi N_j \int_{\nu_j}^{\infty}
\frac{\alpha_{j,\nu}}{h\nu}J_\nu\,\de\nu+\\*+
\sum_{j>i} 4\pi N_j \zav{\frac{N_i}{N_j}}^\ast
\int_{\nu_i}^{\infty}
\frac{\alpha_{i,\nu}}{h\nu}\hzav{\frac{2h\nu^3}{c^2}+J_\nu}
e^{-\frac{h\nu}{kT_\mathrm{e}}}
\,\de\nu-\\*-
\sum_{j<i}4\pi N_i \zav{\frac{N_j}{N_i}}^\ast
\int_{\nu_j}^{\infty}
\frac{\alpha_{j,\nu}}{h\nu}\hzav{\frac{2h\nu^3}{c^2}+J_\nu}
e^{-\frac{h\nu}{kT_\mathrm{e}}}
\,\de\nu-\\*-
\sum_{j>i}4\pi N_i \int_{\nu_i}^{\infty}
\frac{\alpha_{i,\nu}}{h\nu}J_\nu\,\de\nu+
\sum_{j\neq i}N_j \tilde P_{ji} -N_i \sum_{j\neq i}
\tilde P_{ij} =0,
\end{multline}
where we explicitly write rates of direct radiative ionization (I.7a)
and recombination (I.7b),
$\alpha_{i,\nu}$ is the photoionization cross-section from level $i$ with
threshold frequency $\nu_i$ (similarly for $\alpha_{j,\nu}$), $J_\nu$ is the
mean continuum intensity, $T_\mathrm{e}$ is the electron temperature, asterisk
denotes an LTE value and $\tilde P_{ij}$ are rates of all remaining transitions
(collisional ionization, recombination, excitation and deexcitation (I.8) and
radiative bound-bound transitions, equations (I.3), (I.5), (I.14), and the Auger
ionization \eqref{augerclen}).

Introducing the quantity $U_{ji}$ as
\begin{equation}
U_{ji}= n_\mathrm{H}z_\mathrm{atom}
\zav{\frac{N_i}{N_j}}^\ast %\int_{\nu_i}^{\infty}
\alpha_{i,\nu}\frac{2h\nu^3}{c^2}
e^{-\frac{h\nu}{kT_\mathrm{e}}},
\label{defuji}
\end{equation}
the free-bound emissivity has the form of 
\begin{equation}
\eta_\nu=
\sum_{ij,\, j>i}N_j%\int_{\nu_i}^{\infty}
U_{ji},%\,\de\nu.
\label{borohradek}
\end{equation}
where $n_\mathrm{H}$ is the hydrogen number density and $z_\mathrm{atom}$ is
abundance (number densities ratio) of a given atom relative to the hydrogen.
During the process of the solution of the radiative transfer equation we derive
the mean intensity $J_\nu$ from the emissivity \eqref{borohradek} as follows.
Using vector quantities ${\boldsymbol J}_\nu=\zav{J_\nu^1, J_\nu^2, \dots,
J_\nu^{\NR}}^\text{T}$ and ${\boldsymbol \eta}_\nu=\zav{\eta_\nu^1, \eta_\nu^2,
\dots, \eta_\nu^{\NR}}^\text{T}$, where $\NR$ is considered number of depth
points, the radiative transfer equation for a given frequency can be  expressed
in a symbolic form as
\begin{equation}
\label{brandejs}
{\boldsymbol J}_\nu=\Psi_\nu[{\boldsymbol \eta}_\nu],
\end{equation}
where the matrix operator $\Psi_\nu$ represents solution of the radiative
transfer equation. However, the actual process of solution is different, we do
not apply the operator $\Psi_\nu$ on ${\boldsymbol \eta}_\nu$, but the mean
intensity ${\boldsymbol J}_\nu$ is derived as the solution of linear set of
equations
\begin{equation}
\label{bezpravi}
{\boldsymbol \eta}_\nu=\Psi_\nu^{-1}[{\boldsymbol J}_\nu],
\end{equation}
which is in fact the formal solution of the radiative transfer equation. Hence,
during the solution of the radiative transfer equation \eqref{bezpravi} we do
not know an explicit form of $\Psi_\nu$, but we know its inversion
$\Psi_\nu^{-1}$. Since for the acceleration of convergence of statistical
equilibrium equations \eqref{nlte} together with the continuum radiative
transfer equation \eqref{bezpravi} we need to know derivatives ${\partial
J_\nu}/{\partial N_j}$, we consequently need to know the explicit form of
$\Psi_\nu$. However, because for the solution of equations of statistical
equilibrium we use only derivatives of mean intensity with respect of $N_j$ at a
given depth point, we need to know only the diagonal part of the operator
$\Psi_\nu$ (see equation \eqref{borohradek}). This significantly reduces the
necessary computer power. Note that because we use LAPACK package ({\tt
http://www.cs.colorado.edu/\~{}lapack}, Anderson et al. \citeyear{lapack}) for
the solution of radiative transfer equation \eqref{bezpravi}, which is based on
LU decomposition, we can easily calculate diagonal elements of $\Psi_\nu$ (see
Appendix~\ref{usti}). Finally, derivatives of mean intensities inserted into the
statistical equilibrium equations have at the depth point $\IR$ the approximate
form of
\begin{equation}
\label{trebova}
\left.\frac{\partial J_\nu}{\partial N_i}\right|_{\IR}\approx
\Psi_{\nu,\IR,\IR}U_{ij},
\end{equation}
where $\Psi_{\nu,\IR,\IR}$ is the
corresponding diagonal element of $\Psi_\nu$ at a depth point $\IR$
and $U_{ij}$ was defined in equation~\eqref{defuji}.

During the iterative solution of statistical equilibrium equations we solve for
the corrections $\delta N_i$ to the actual relative occupation numbers $N_i$.
These corrections are calculated using equation~(I.22). Here we also add
corrections due to the dependence of the mean intensity on $N_i$
(equation~\ref{trebova}). Hence, instead of equations~(I.22) we solve (compare
with equation~\ref{poradtrebova})
\begin{multline}
\label{asibezdomovec}
\sum_{j<i}4\pi N_j \int_{\nu_j}^{\infty}
\frac{\alpha_{j,\nu}}{h\nu}
\Psi_{\nu,\IR,\IR}^\ast U_{ij}\delta N_i\,\de\nu
+\\*+
\sum_{j>i} 4\pi N_j \zav{\frac{N_i}{N_j}}^\ast
\int_{\nu_i}^{\infty}
\frac{\alpha_{i,\nu}}{h\nu}\Psi_{\nu,\IR,\IR}^\ast U_{ji}\delta N_j
e^{-\frac{h\nu}{kT_\mathrm{e}}}
\,\de\nu -\\*-
\sum_{j<i}4\pi N_i \zav{\frac{N_j}{N_i}}^\ast
\int_{\nu_j}^{\infty}
\frac{\alpha_{j,\nu}}{h\nu}\Psi_{\nu,\IR,\IR}^\ast U_{ij}\delta N_i
e^{-\frac{h\nu}{kT_\mathrm{e}}}
\,\de\nu-
\\*-
\sum_{j>i}4\pi N_i \int_{\nu_i}^{\infty}
\frac{\alpha_{i,\nu}}{h\nu}\Psi_{\nu,\IR,\IR}^\ast U_{ji} \delta N_j
\,\de\nu
+\\+\sum_{j\neq i}\hzav{N_j\frac{\partial P_{ji}}{\partial N_i}\delta N_i+
      \zav{P_{ji}+ N_j\frac{\partial P_{ji}}{\partial N_j}} \delta N_j}-
\\*
     -\sum_{j\neq i}\hzav{\zav{P_{ij}+
      N_i\frac{\partial P_{ij}}{\partial N_i}}\delta N_i+
      N_i\frac{\partial P_{ij}}{\partial N_j} \delta N_j}=0,
\end{multline}
where first four sums represent the term of accelerated lambda iterations and
the meaning of other terms is the same as in Paper~I.

Finally, for the calculation of continuum mean intensities $J_\nu$ (see
equation~\ref{bezpravi}) we use the formal solution (i.e.~for given opacity and
emissivity) of the momentum form of the radiative transfer equation with
inclusion of sphericity factors (Auer \citeyear{auer}).

To accelerate the convergence even more, we have also included the Ng
acceleration (Ng \citeyear{ng}, see also \citealt*{nrtauer} or Hubeny
\citeyear{hubtueb}).

\section{Studied stars}

This paper studies the influence of X-rays on the basic parameters of
radiatively driven stellar winds.
However, to avoid using of ad hoc stellar parameters, we rather decided
to choose a set of parameters corresponding to real stars with basic
parameters already known to some degree of accuracy.

\subsection{Stellar parameters}

Our study is based on O stars with effective temperatures
$\Teff\lesssim40\,000\,$K, which were detected by {\em ROSAT} satellite as X-ray
sources (Bergh\"ofer et al. \citeyear{rosat}, hereafter BSC). They were selected
out of stars studied by Repolust et al. (\citeyear{rep}, hereafter R04), Markova
et al. (\citeyear{upice}, hereafter M04), and in the Paper~I. To enlarge our
dataset we also included such stars from Martins et al. (\citeyear{martclump},
hereafter \citetalias{martclump}) sample, for which measured X-ray fluxes are
available in the literature. All but one of these stars exhibit the so-called
"weak wind problem", i.e., their theoretically predicted mass-loss rates are
significantly higher than those derived from observations.

\begin{table}
\caption{Stellar parameters of selected O stars. Stars exhibiting the "weak wind
problem" appear in the bottom part below the horizontal line.
Spectral types are taken from the Simbad database.
}
\label{obhvezpar}
\centering
\begin{tabular}{rrccrcc}
\hline
\multicolumn{1}{c}{Star} & \multicolumn{1}{c}{HD} & Sp. & ${R_{*}}$ & $M$ &
$\Teff$ & Source \\
& \multicolumn{1}{c}{number}& type & $[\text{R}_{\odot}]$ &$[\text{M}_{\odot}]$ &  $[\text{K}] $ \\
\hline
$\xi$ Per      &  $24912$ & O7.5IIIe & $14.0$ & $36$ & $35\,000$ & R04 \\%!
$\alpha$ Cam   &  $30614$ & O9.5Iae & $27.6$ & $43$ & $30\,900$ & LSL \\
$\lambda$ Ori A&  $36861$ & O8 III   & $12.3$ & $30$ & $36\,000$ & LSL \\%!
               &  $54662$ & O7III    & $11.9$ & $38$ & $38\,600$ & M04 \\
	       &  $93204$ & O5V      & $11.9$ & $41$ & $40\,000$ & M05 \\
$\zeta$ Oph    & $149757$ & O9V      &  $8.9$ & $21$ & $32\,000$ & R04 \\
63 Oph         & $162978$ & O8III    & $16.0$ & $40$ & $37\,100$ & LSL \\%?
68 Cyg         & $203064$ & O8e      & $15.7$ & $38$ & $34\,500$ & R04 \\
19 Cep         & $209975$ & O9Ib     & $22.9$ & $47$ & $32\,000$ & R04 \\%!
$\lambda$  Cep & $210839$ & O6Iab    & $19.6$ & $51$ & $38\,200$ & LSL \\
\hline AE Aur  & $34078$  & O9.5Ve   & $ 7.5$ & $20$ & $33\,000$ & M05 \\
$\mu$ Col      & $38666$  & O9.5V    & $ 6.6$ & $19$ & $33\,000$ & M05 \\
               & $42088$  & O6.5V    & $ 9.6$ & $31$ & $38\,000$ & M05 \\
               & $46202$  & O9V      & $ 8.4$ & $21$ & $33\,000$ & M05 \\
\hline
\end{tabular}
\end{table}

Adopted parameters of studied O stars are given in Table~\ref{obhvezpar}.
Effective temperatures and radii are taken from R04, M04,
\citetalias{martclump}, and Lamers et al. (\citeyear{lsl}, hereafter LSL).
Parameters derived by R04, M04, and M05 were obtained using blanketed model
atmospheres, i.e., they are more reliable than the older ones. Stellar masses
were obtained using evolutionary tracks either by us (using tracks calculated by
Schaller et al.~\citeyear{salek}) or by LSL or M05. The use of the evolutionary
masses instead of the spectroscopic ones may cause a systematic shift due to the
well-known discrepancy between these masses \citep[e.g.][]{hekuku}. However, for
many stars from our sample these masses are nearly the same. 

\subsection{Wind parameters derived from observations}
\label{pozor}

Wind parameters of studied stars (both derived from observations and predicted
ones) are given in Table~\ref{obvitpar}.

X-ray luminosities, which are assumed to originate in the wind, were taken from
\citetalias{rosat} with an exception of stars HD~42088, HD~46202, and HD~93204,
for which the X-ray luminosities were taken from \citet{pekarna} and
\citet{xeva}.

Because there is still no broad consensus about the influence of clumping on
mass-loss rates derived from observations, we used mass-loss estimations from
\citet [hereafter P06] {pulchuch}, which were regarded as upper limits with
respect to clumping, supplemented by results of M05. Although M05 derived
mass-loss rates of some stars with inclusion of clumping, for most stars
selected by us from this sample the clumping factor was set by them to one,
because these stars exhibit the "weak wind problem". The only exception is the
star HD~93204, for which M05 provide mass-loss rate with clumping taken into the
account. However, to keep our sample more compact, we used mass-loss rate
uncorrected for clumping (calculated as $\dot M/\sqrt{f_\infty}$, where
$f_\infty$ is the clumping factor in the outer wind derived by M05). Because
\citetalias{pulchuch} concluded that mass-loss rates derived from radio data are
less influenced by clumping, for the star HD~149757 we adopted such rates
derived by \citet{lamlei}.

Terminal wind velocities were taken from LSL, Puls et al. (\citeyear{pulmoc}),
and M04 (the uncertainties were either taken from LSL or calculated assuming
10\% errors as suggested by Puls et al. \citeyear{pulmoc}). For stars that
exhibit the "weak wind problem" the terminal velocities derived from
observations may be just lower limits, similarly as in \citet{martin}.
Consequently, we did not consider the terminal velocities of these stars in the
following analysis.

\newcommand\mez{\hspace{1mm}}
\newcommand\mmez{\hspace{0mm}}

\begin{table*}
\centering
\caption{Wind and X-ray parameters of studied O stars. Wind parameters derived
from observation are described in Sect.~\ref{pozor}. Predicted wind parameters
were derived using our NLTE models with different properties of X-ray sources
(models with $\fx=0$, $\fx=0.02$ and $u_\text{rel}=0.3$).}
\label{obvitpar}
\begin{tabular}{rc@{\mez}c@{\mez}cc@{\mez}cc@{\mez}c@{\mez}c}
\hline
\multicolumn{1}{c}{HD} & 
\multicolumn{3}{c}{Observed} &
\multicolumn{2}{c}{Predicted no X-rays}&
\multicolumn{3}{c}{Predicted -- $\fx=0.02$}\\
\multicolumn{1}{c}{number}&
$\log L_\text{X}$ & \dmdt & $v_\infty$ &
$\dmdt$ & $v_\infty$ & $\log L_\text{X}$ & $\dmdt$ & $v_\infty$\\
& [CGS] & {$[\msr]$}& [\kms] &
$[\msr]$ & [\kms] & [CGS] & $[\msr]$ & [\kms]\\
\hline
 $24912$ & 31.91 & $1.2\times10^{-6}$  & $2450$ & $4.4\times10^{-7}$ & $2270$ & 31.71& $4.4\times10^{-7}$ & $2750$\\
 $30614$ & 32.24 & $1.5\times10^{-6}$  & $1500$ & $1.5\times10^{-6}$ & $1950$ & 32.27 & $1.4\times10^{-6}$ & $2290$ \\
 $36861$ & 32.59 & $4\times10^{-7}$    & $2200$ & $4.8\times10^{-7}$ & $2150$ & 31.80& $4.8\times10^{-7}$ & $2500$  \\
 $54662$ & 32.34 &                    & $2450$ & $7.9\times10^{-7}$ & $2190$ & 32.10 & $7.9\times10^{-7}$ & $2480$ \\
 $93204$ & 32.07 & $5.6\times10^{-7}$  & $2900$ & $1.3\times10^{-6}$ & $2290$ & 32.40 & $1.3\times10^{-6}$ & $2400$ \\
$149757$ & 31.14 & $3.9\times10^{-8}$  & $1550$ & $4.7\times10^{-8}$ & $2040$ & 30.23 & $4.7\times10^{-8}$ & $2280$ \\
$162978$ & 32.95 &                    & $2200$ & $2.0\times10^{-6}$ & $2040$ & 32.50 & $1.9\times10^{-6}$ & $2190$ \\
$203064$ & 31.76 & $1.1\times10^{-6}$  & $2550$ & $5.7\times10^{-7}$ & $2080$ & 31.87 &$5.8\times10^{-7}$ & $2590$ \\
$209975$ & 32.47 & $1.2\times10^{-6}$  & $2050$ & $8.4\times10^{-7}$ & $2430$ & 31.99 & $8.4\times10^{-7}$ & $2900$ \\
$210839$ & 32.12 & $3.0\times10^{-6}$  & $2200$ & $6.1\times10^{-6}$ & $1990$ & 32.95 & $6.0\times10^{-6}$ & $1910$ \\
\hline
 $34078$ & 31.32 & $3.2\times10^{-10}$ &  $800 $    & $1.4\times10^{-8}$ & $2950$ & $29.34$ & $1.4\times10^{-8}$ & $3490$ \\
 $38666$ & 31.80 & $3.2\times10^{-10}$ &  $1200$    & $7.9\times10^{-9}$ & $4380$ & $28.78$ & $8.0\times10^{ -9}$ & $4480$ \\
 $42088$ & 32.38 & $1\times10^{-8}   $ &  $1900$    & $3.1\times10^{-7}$ & $2180$ & $31.63$ & $3.2\times10^{-7}$ & $2680$ \\
 $46202$ & 32.40 & $1.3\times10^{-9} $ &  $1200$      & $2.3\times10^{-8}$ & $1900$ & 29.76 & $2.4\times10^{-8}$ & $2890$ \\
\hline
\end{tabular}
\end{table*}

\section{The emergent X-rays}

\subsection{The $L_\text{X}/L$ relation}
\label{kaplxskal}

From the analytic considerations \citet[hereafter \citetalias{oskal}]{oskal}
showed that the optically thin X-ray luminosity depends on the square of the
mass-loss rate $L_\text{X}\sim(\dot M/v_\infty)^2$, whereas the X-ray luminosity
of the optically thick wind scales linearly with the mass-loss rate
$L_\text{X}\sim\dot M/v_\infty$. A slightly different form of these relations
can be expected due to the assumed dependence of the shock temperature
$T_\text{X}$ on the wind terminal velocity. Consequently, in the following we
use simpler relations $L_\text{X}\sim\dot M^2$ for the optically thin wind in
the X-ray region and $L_\text{X}\sim\dot M$ for the optically thick wind. From
the relation $\dot M\sim L^{1/\alpha'}$ \citep{kupul}, where
$\alpha'=\alpha-\delta$ and $\alpha$ and $\delta$ are usual CAK force
multipliers, the X-ray luminosity of optically thick wind is predicted to be
proportional to $L_\text{X}\sim L^{1/\alpha'}$ ($\alpha'\approx0.5$). 

However, from the observations a slightly different trend emerges, because the
observed X-ray luminosity $L_\text{X}$ is roughly linearly proportional to the
total luminosity  $L$ \citep[e.g.][]{chleba,rosat,sane}. The origin of the
difference is not clear. \citetalias{oskal} suggested that the observed relation
can be reproduced assuming that \fx\ decreases with radius.

\begin{figure}
\centering
\resizebox{0.9\hsize}{!}{\includegraphics{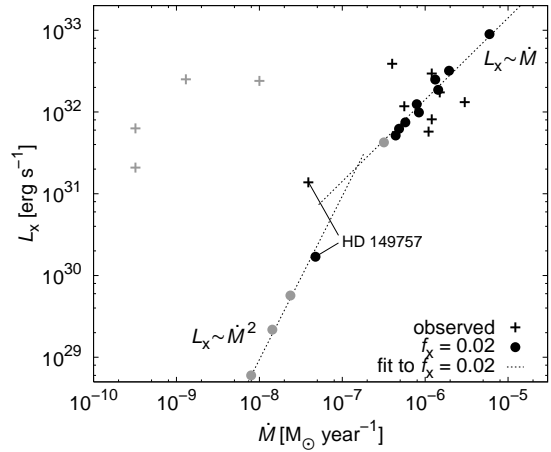}}
\caption{The relation between $L_\text{X}$ and $\dot M$ plotted using wind
parameters derived from observations and from theoretical models ($\fx=0.02$).
The relations $L_\text{X}\sim\dot M$ for stars with optically thick winds in the
outer regions (stars with large mass-loss rates) and $L_\text{X}\sim\dot M^2$,
valid for stars with optically thin winds \citepalias{oskal} are also plotted in
the graph (dotted lines). Gray symbols denote values for stars exhibiting "weak
wind problem".}
\label{lxdmvn}
\end{figure}

\begin{figure}
\centering
\resizebox{0.9\hsize}{!}{\includegraphics{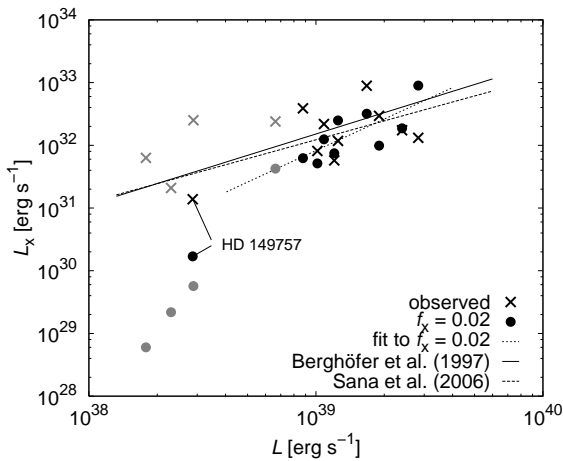}}
\caption{The relation between the X-ray luminosity $L_\text{X}$ and the total
luminosity $L$ for studied stars calculated assuming $\fx=0.02$.
Overplotted are the average observed
relations derived by \citet{rosatvel}, \protect\citet{sane}, observed values for
studied stars, and the linear fit to the theoretical expectations
(only for stars with thick winds). Gray symbols denote values for stars
exhibiting "weak wind problem".}
\label{lxlbolhv}
\end{figure}

To understand the origin of $L_\text{X}/L$ relation we calculated wind models
with a fixed filling factor. To be specific, we chose $\fx=0.02$. In agreement
with the calculations of \citetalias{oskal}, the X-ray luminosity of individual
stars depends mostly on the wind mass-loss rate (see Fig.~\ref{lxdmvn}). Stars
with large mass-loss rates $\dot M\gtrsim10^{-7}\,\msr$ have an optically thick
wind in the X-ray domain (between the outermost model point and the region where
most of the X-rays are generated) and the X-ray luminosity of theoretical models
calculated with a fixed \fx\ depends on $\dot M$ linearly. On the other hand,
stars with low mass-loss rates $\dot M\lesssim10^{-7}\,\msr$ have optically thin
wind and their theoretical X-ray luminosity depends on $\dot M^2$.

The $L_\text{X}-\dot M$ relation plotted using wind parameters derived from
observations (Fig.~\ref{lxdmvn}) is different. The correlation between the
observed mass-loss rate and X-ray luminosity is less apparent. Moreover, the
observed X-ray luminosities for stars with low mass-loss rates
$\dmdt\lesssim10^{-7}\msr$ lie considerably above the theoretical expectations.
These stars have nearly the same $L_\text{X}$ as stars with high mass-loss
rates.

The predicted dependence of $L_\text{X}$ on $L$ displayed in Fig.~\ref{lxlbolhv}
reflects the dependence of $L_\text{X}$ on $\dot M$. For stars with dense winds,
i.e., for stars with large luminosities
($L\gtrsim5\cdot10^{38}\,\text{erg}\,\text{s}^{-1}$) the X-ray luminosity scales
as $\dot M$, and due to the dependence of $\dot M$ on $L$, $L_\text{X}$ depends
mostly on the stellar luminosity as $L_\text{X} \sim L^{1.7}$. This is in
agreement with scaling $L_\text{X} \sim L^{1/\alpha'}$ where $\alpha'\approx
0.6$. For stars with thick winds these models roughly recover the observed
$L_\text{X}/L$ relation. Remaining difference between predicted slope of
$L_\text{X}/L$ relation and the observed one may be either due the radial
dependence of the filling factor \citepalias[as proposed by][]{oskal}, or due to
the dependence of the filling factor on wind density (Sect.~\ref{hustopln}).
Higher scatter of the observed X-ray luminosities may be partly caused by poorly
known distances \citep[cf. with the results of][]{sane}.

For stars with lower luminosities,
$L\lesssim5\cdot10^{38}\,\text{erg}\,\text{s}^{-1}$, there is an abrupt change
in $L_\text{X}/L$ relation derived from our models with fixed \fx. These stars
have optically thin wind, and their X-ray luminosities scale as
$L_\text{X}\sim\dot M^2$. Because the mass-loss rate depends mainly on the
stellar luminosity, the $L_\text{X}/L$ relationship derived from theoretical
models consequently steepens. However, the observed X-ray luminosities are much
higher than the predicted ones. The observational $L_\text{X}/L$ relation is
almost the same for both groups of stars with optically thick and thin winds.
Because in the case of optically thin winds we observe basically all X-rays
emitted, our description of shock properties of these stars is likely
oversimplified. Other possibilities, like different heating mechanism
\cite[e.g., wind frictional heating,][]{kapusty} are also not ruled out.

The enhanced X-ray activity may not be limited only to the stars exhibiting the
``weak wind problem". 
%Kr1: Observations show
From observations of \citet{rosatvel} follows
that also many other late O main
sequence stars show X-ray luminosity corresponding to the stars with thick
winds.

\subsection{The radial distribution of X-ray emissivity}

\begin{figure}
\centering
\resizebox{0.9\hsize}{!}{\includegraphics{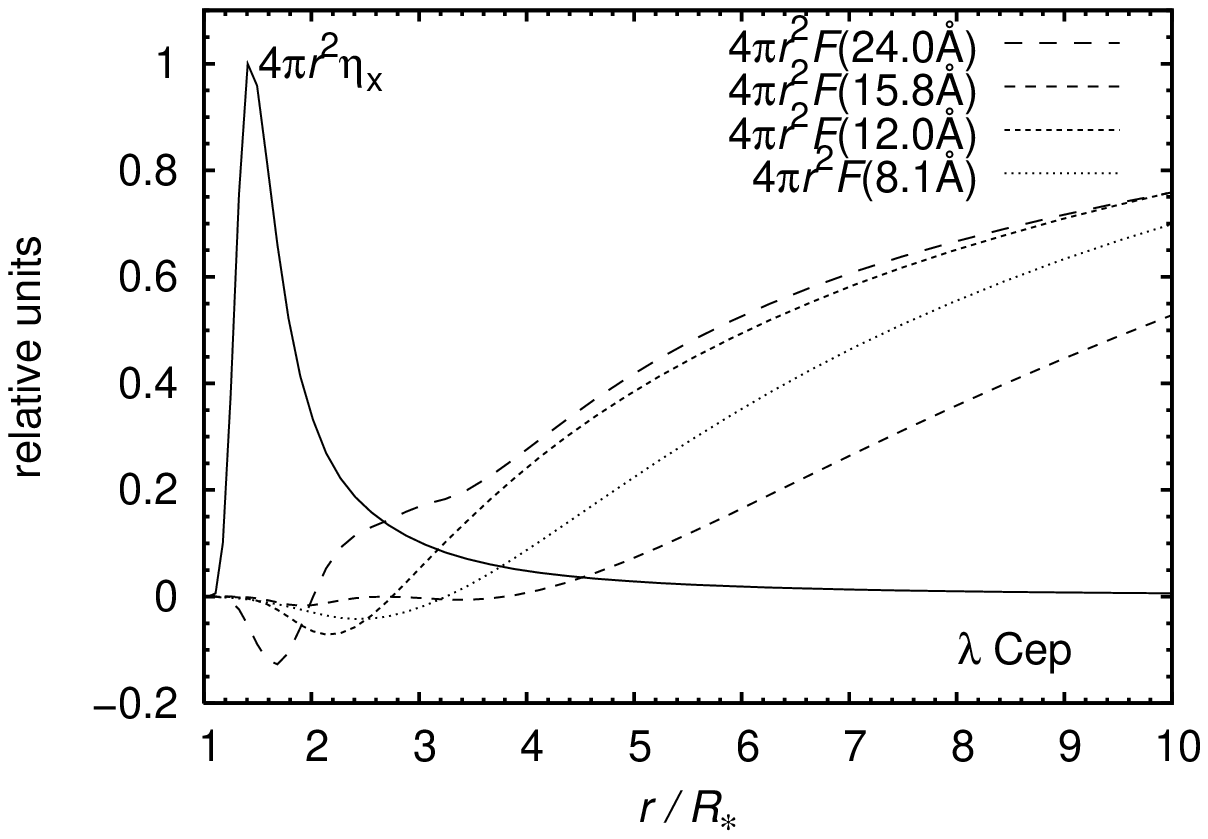}}
\resizebox{0.9\hsize}{!}{\includegraphics{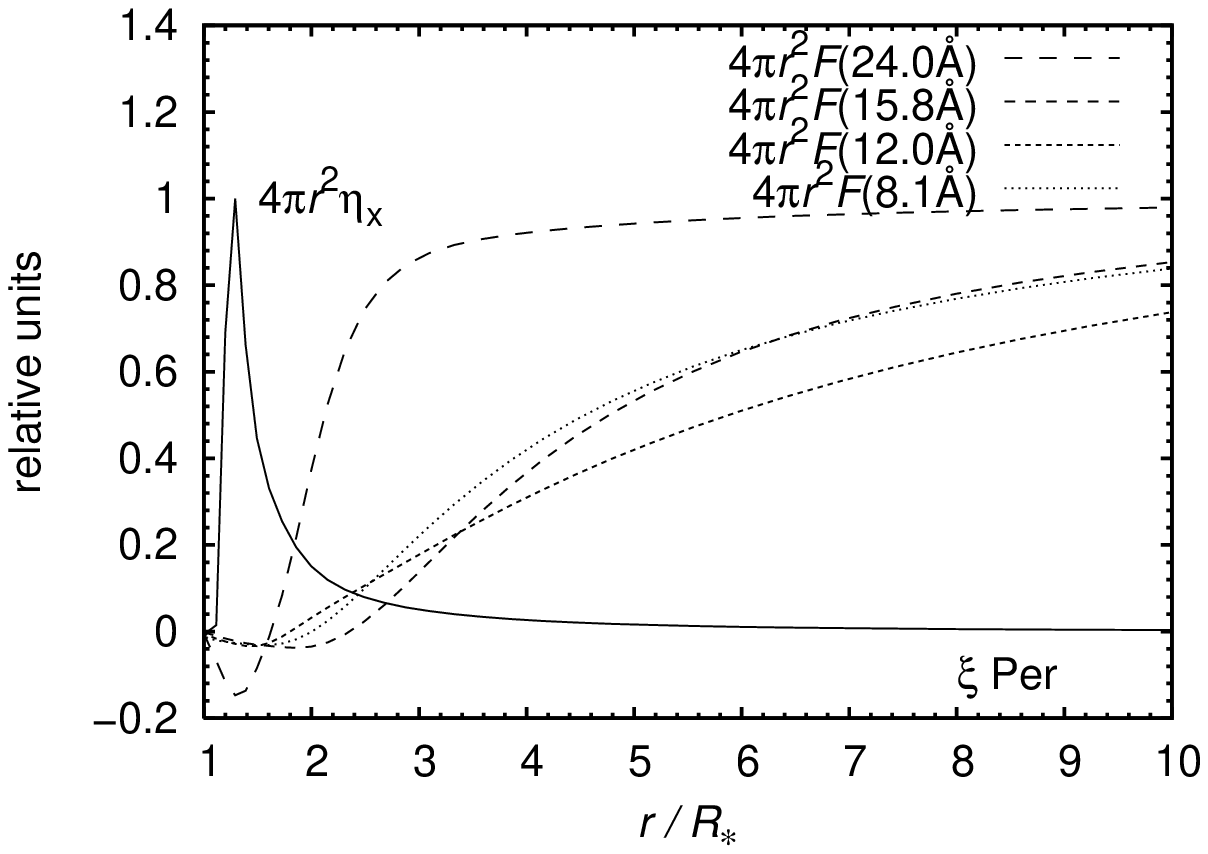}}
\caption{The radial dependence of the the total shock emissivity (integrated
over frequencies and plotted as $4\pi r^2\eta_\text{X}$) and the monochromatic
luminosity $4\pi r^2 F(\lambda)$ for $\fx=0.02$ for stars $\lambda$~Cep and
$\xi$~Per. The quantities are plotted relatively to their maximum values.
%Kr1:
Note that the integrated emissivity has maximum below the radius at which
the
%Ku1:
X-ray
luminosities reach the observed value. This indicates that most of X-rays
emitted in the wind do not escape.}
\label{etaxr}
\end{figure}

Most of X-rays in our models are generated close to the star, as shown in the
plot of radial variations of X-ray emissivity Fig.~\ref{etaxr}. Due to assumed
dependence of the shock temperature on wind velocity (via equations \eqref{tx},
\eqref{urel}), the X-ray emissivity is increasing function of radius close to
the star. On the other hand, we assume that the X-ray emissivity depends on the
square of the wind density (see equation \eqref{etax}), consequently the
emissivity decreases in the outer regions.

For stars with optically thick wind only a very small part of generated X-rays
finally escapes the wind and may reach a distant observer, as can be seen from
the difference between the radii at which $4\pi r^2\eta_\text{X}$ has maximum
and at which the monochromatic luminosities approach the terminal value in
Fig.~\ref{etaxr}. Due to the continuum X-ray absorption also the X-ray
luminosity of optically thick wind depends less steeply on the wind mass-loss
rate than the X-ray luminosity of optically thin wind (see Fig.~\ref{lxdmvn} and
OC). The X-ray flux is not always monotonically increasing, and in some regions
it is even negative due to the transfer of part of emitted X-rays towards the
stellar surface. 

\begin{figure}
\centering
\resizebox{0.7\hsize}{!}{\includegraphics{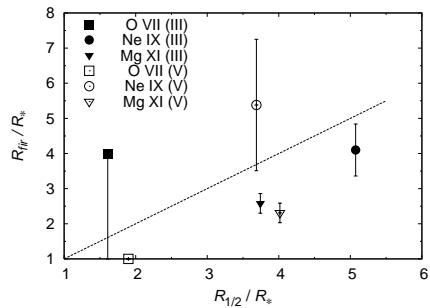}}
\caption{The relation between the radius \rfir\ of the formation of $fir$ lines
of He-like ions derived from observations by \citet{walca} and the radius
$R_{1/2}$ at which the monochromatic luminosity is equal to
half of its terminal value (for $\fx=0.02$). The radii derived for individual
lines for giants (III) and main-sequence stars (V) are plotted using different
symbols. Dashed line denotes one to one relation.}
\label{rfirrf}
\end{figure}

In the outer regions the monochromatic luminosities $4\pi r^2 F(\lambda)$ in
Fig.~\ref{etaxr} rise and reach the value measured by a distant observer as the
wind becomes optically thin. \citet{walca} showed that there is a correlation
between the radius \rfir\ of the formation of forbidden intercombination
resonance ($fir$) lines of He-like ions derived from observations, and the
radius at which the continuum optical depth at associated wavelength is unity.
Note, however, that the formation radius of the observed X-rays depends not only
on opacity (optical depth), as \citet{walca} stressed, but naturally also on
emissivity. Consequently, we argue that the relationship found by \citet{walca}
is more precisely manifested as a correlation between \rfir\ and the radius
$R_{1/2}$ at which the monochromatic luminosity at the associated wavelength is
equal to half of its terminal value (Fig.~\ref{rfirrf}).

\subsection{The hardness ratio}

\begin{figure}
\centering
\resizebox{0.8\hsize}{!}{\includegraphics{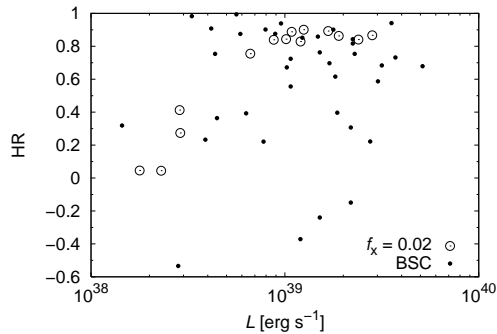}}
\caption{The comparison between observed (derived by \protect\citetalias{rosat}
for O stars) and model hardness ratio defined by equation~\eqref{harddef}.}
\label{hardobr}
\end{figure}

To test whether the derived spectral energy distribution of X-rays is
at least roughly correct, we compare the predicted and observed hardness
ratio, introduced by BSC as
\begin{equation}
\label{harddef}
\text{HR}=(\text{H}-\text{S})/(\text{H}+\text{S}),
\end{equation}
where H and S are the
integrated
fluxes in the hard ($0.5-2.0\,$keV) and soft
($0.1-0.4\,$keV) energy bands, respectively
%Kr1: . The comparison in Fig.~\ref{hardobr} shows that
%our models predict slightly harder X-rays than observed.
(see Fig.~\ref{hardobr}).
The model hardness ratio is nearly constant for stars
with large luminosities. It decreases for stars with lower luminosities, because
the stellar wind of these stars is less opaque and regions close to the star
(where X-rays with lower energies are generated) become visible.
%Kr1:
The observed values for many stars
%Ku1: are however
are, however,
lower than the predicted ones and display
%Ku1: a
%
larger scatter. In the frame of our model this means that the
observations cannot be explained just by a single value of $u_\text{rel}$.

\section{Influence of X-rays on the ionization state}

\begin{figure}
\centering
\resizebox{0.9\hsize}{!}{\includegraphics{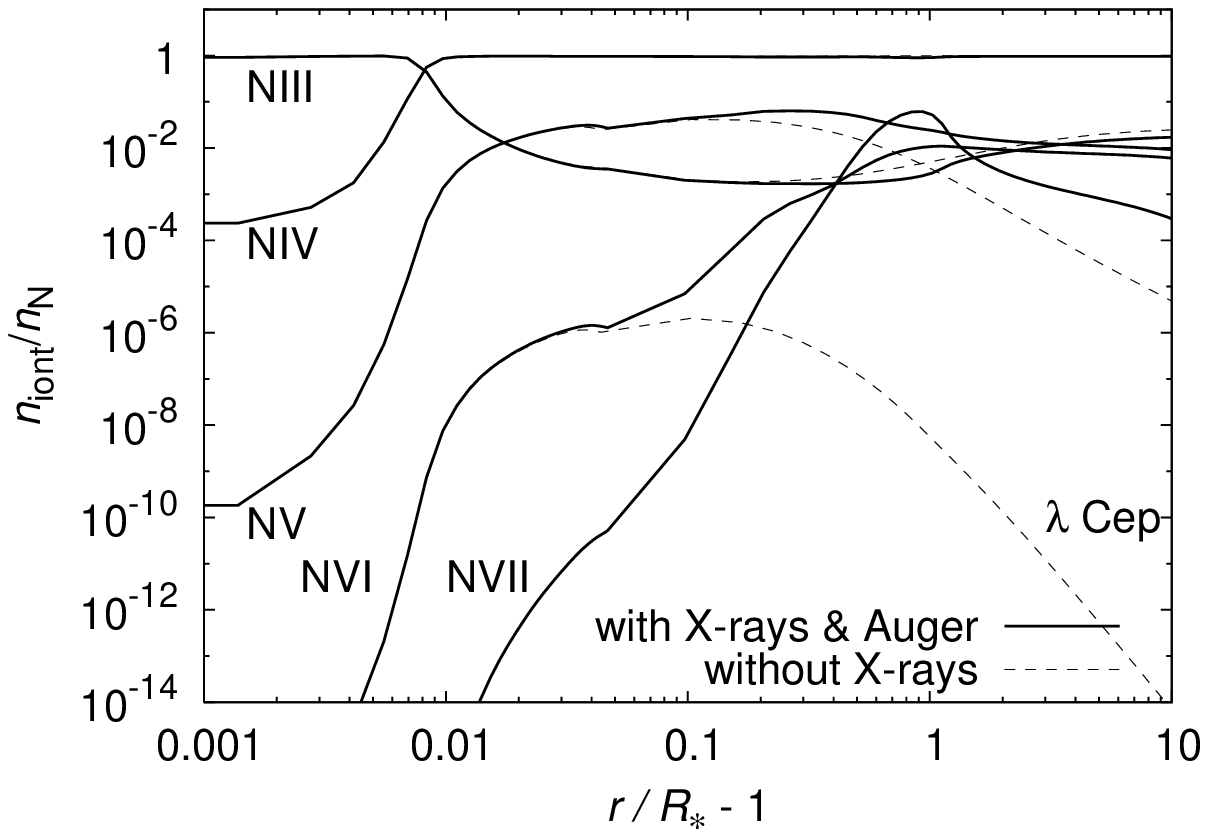}}
\resizebox{0.9\hsize}{!}{\includegraphics{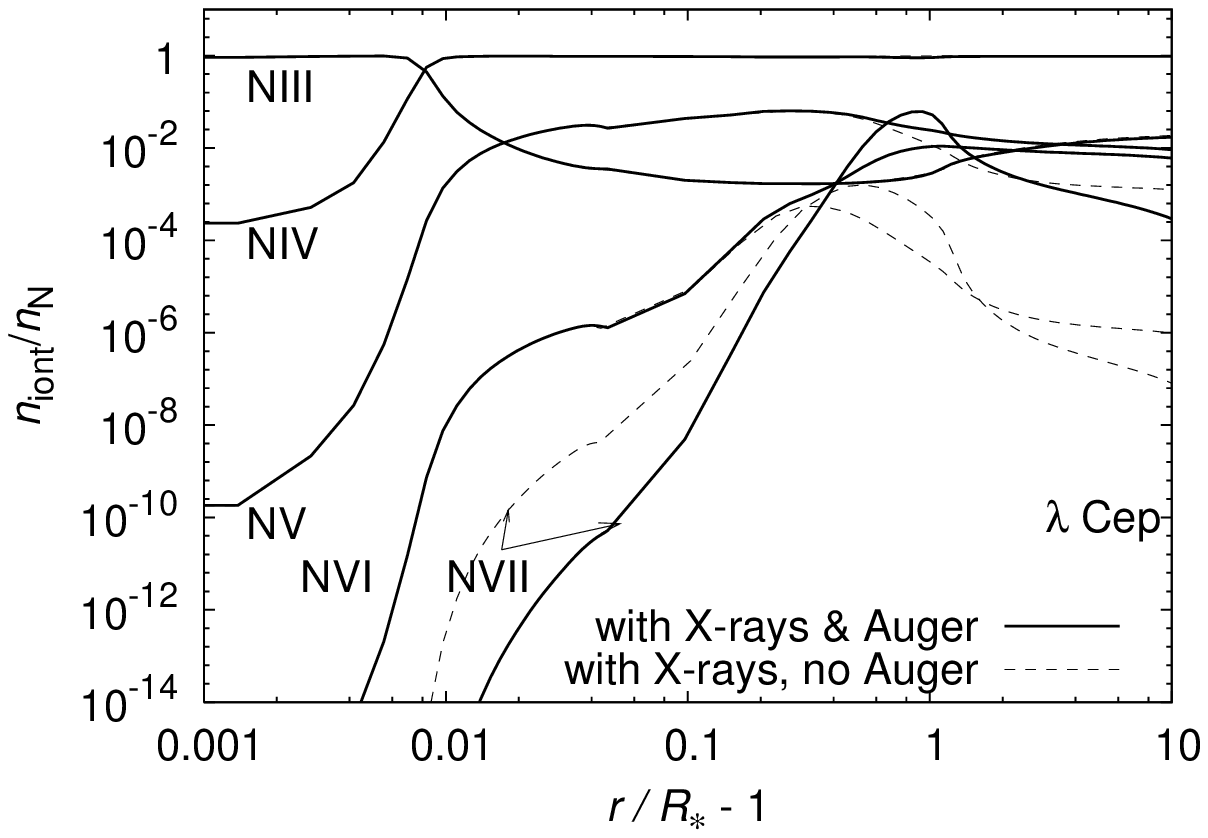}} 
\caption{Radial variation of nitrogen ionization fractions in the wind model of
$\lambda$~Cep for $\fx=0.0028$ that fits the observed $L_\text{X}$.
{\em Top panel:} Comparison of models calculated with
additional X-ray sources (solid lines) and without these sources (dashed lines).
{\em Bottom panel:} Comparison of models with additional X-ray sources
calculated with inclusion of Auger processes (solid lines) and neglecting the
Auger processes (dashed lines). For better readability,
the ionization fraction of \ion{N}{ii} is not plotted.}
\label{210839ion} 
\end{figure}

Although not directly observable, ionization structure of the wind
gives us a useful direct insight to the effect of different  radiation
processes in the stellar atmosphere and wind.
In our models the X-rays affect wind ionization structure via direct and Auger
ionization (MacFarlane et al.~\citeyear{macown}, Pauldrach et
al.~\citeyear{lepsipasam}).

\subsection{Radial variations of the ionization fractions}

As an example we discuss the variations of nitrogen
ionization in the case of the star $\lambda$~Cep (HD~210839). To understand the
influence of individual ionization mechanisms on the wind ionization structure
we also calculated models without additional X-ray sources and models which take
these sources into account, but which neglect Auger processes (see
Figs.~\ref{210839ion} for the plot of the variations of nitrogen ionization
fractions with the radius).

The ionization fractions of most abundant ions (\ion{N}{iii}, \ion{N}{iv}) close
to the stellar surface are not significantly influenced by the additional
sources of X-rays. This region is optically thick for the X-ray radiation and
only a very tiny amount of X-rays emitted in the highly supersonic wind parts
may penetrate close to the stellar surface and modify the ionization fractions
of minor higher ions there (in our case \ion{N}{vi} and \ion{N}{vii}). With
increasing radius (for $r/R_\ast-1\gtrsim0.1$) the wind becomes more ionised due
to lower wind density, and X-rays start to play a more important role in the
ionization balance. As a result of our assumptions, the Auger ionization
significantly influences the ionization balance only in the outer wind regions.
This is caused by the fact that more energetic X-rays emitted in the wind with
sufficiently high velocity are necessary for the Auger ionization than for
direct ionization.

\begin{figure}
\centering
\resizebox{0.9\hsize}{!}{\includegraphics{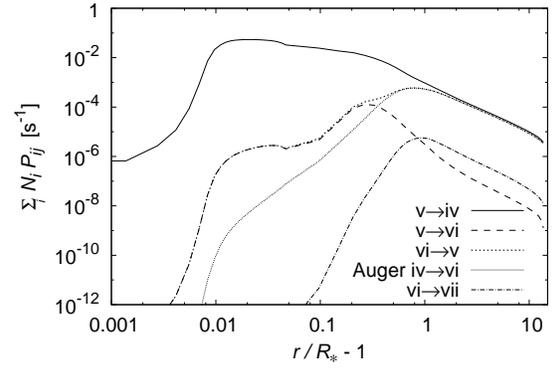}}
\caption{The radial dependence of selected sums of radiative
ionization/recombination rates $\sum_i N_iP_{ij}$ in the model of $\lambda$~Cep
($\fx=0.0028$).
Rates are summed over all transitions between corresponding ions.
Close to the star, the ionization fraction of \ion{N}{v} is given by the
radiative ionization balance with \ion{N}{iv}. In the outer regions also the
recombination from \ion{N}{vi} influences the fraction of \ion{N}{v}.
Consequently, \ion{N}{v} partly occurs due a two-step process (Auger ionization
of \ion{N}{iv} followed by the radiative recombination).}
\label{210839cet}
\end{figure}

The ionization fraction of \ion{N}{v} is increased due to direct ionization from
\ion{N}{iv} and is influenced also by ionization to and recombination from
\ion{N}{vi}. The ionization fraction of \ion{N}{vi} is increased by both Auger
and direct ionization. \ion{N}{vii} is generated due to the direct ionization of
\ion{N}{vi} (see Fig.~\ref{210839cet}).

The result that for $r/R_\ast-1\lesssim0.5$ the ionization fraction of
\ion{N}{vii} is higher in the case of neglected Auger processes than in the
model with these processes may seem paradoxical, but it has a simple explanation.
In the case of neglected Auger processes the stellar wind is less opaque in the
X-ray region and, consequently, the intensity of X-ray radiation is higher.
Because \ion{N}{vii} is created via direct ionization of \ion{N}{vi}, the
ionization fraction of \ion{N}{vii} is higher.

\subsection{Variations with the effective temperature}

\begin{figure*}
\centering
\resizebox{0.33\hsize}{!}{\includegraphics{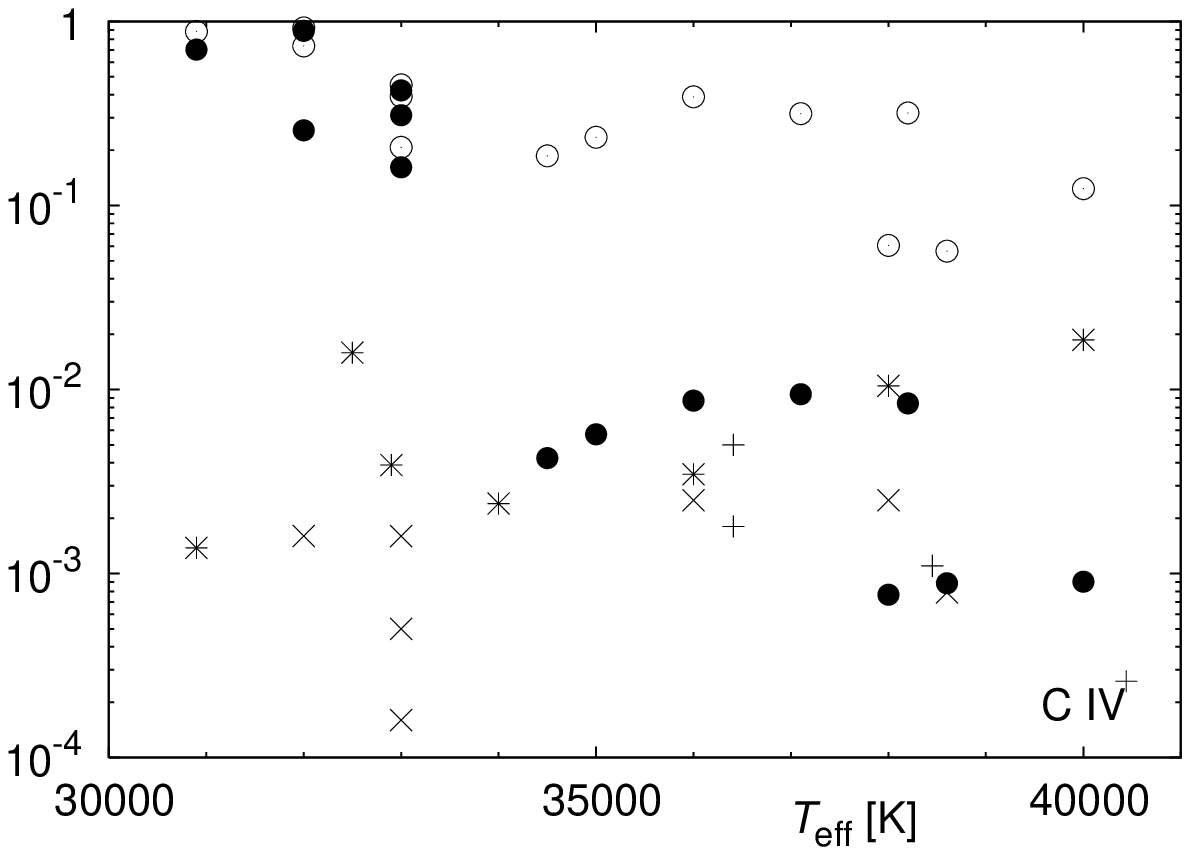}}
\resizebox{0.33\hsize}{!}{\includegraphics{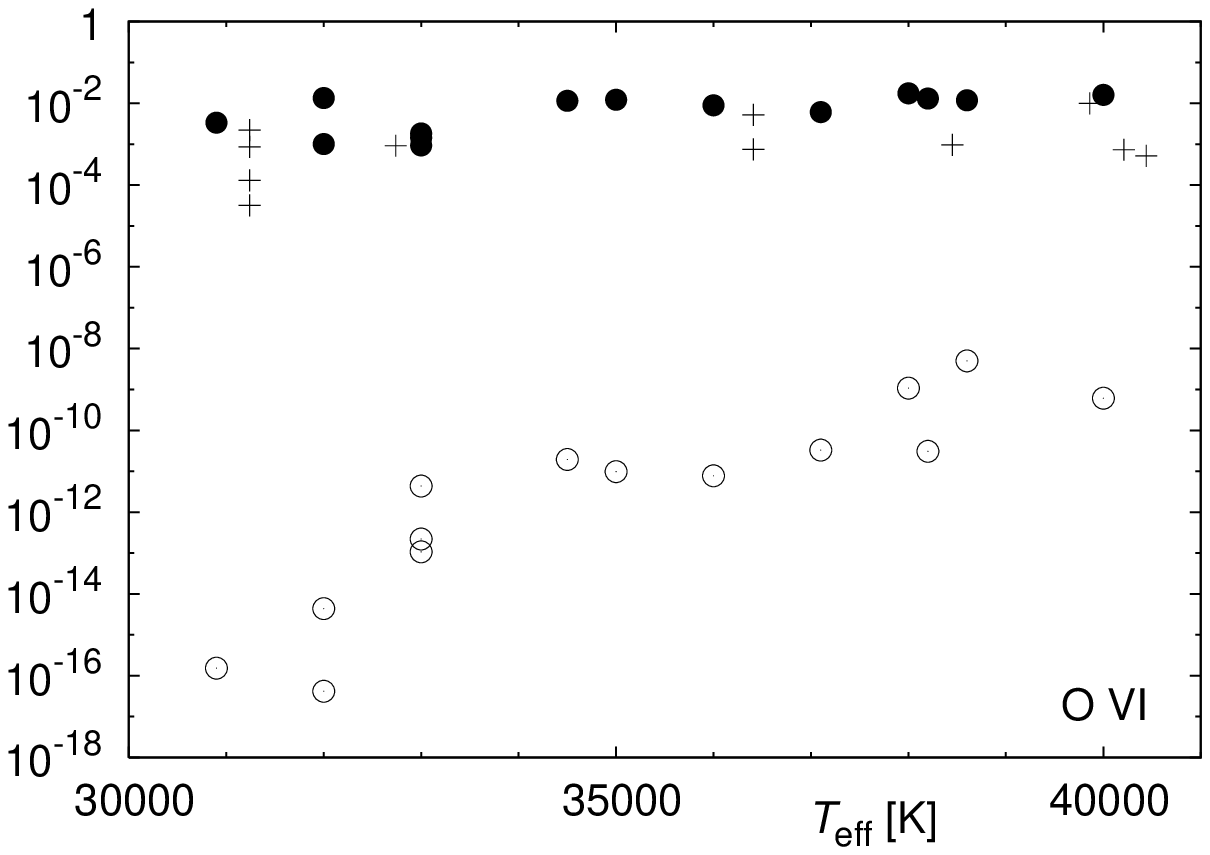}}
\resizebox{0.33\hsize}{!}{\includegraphics{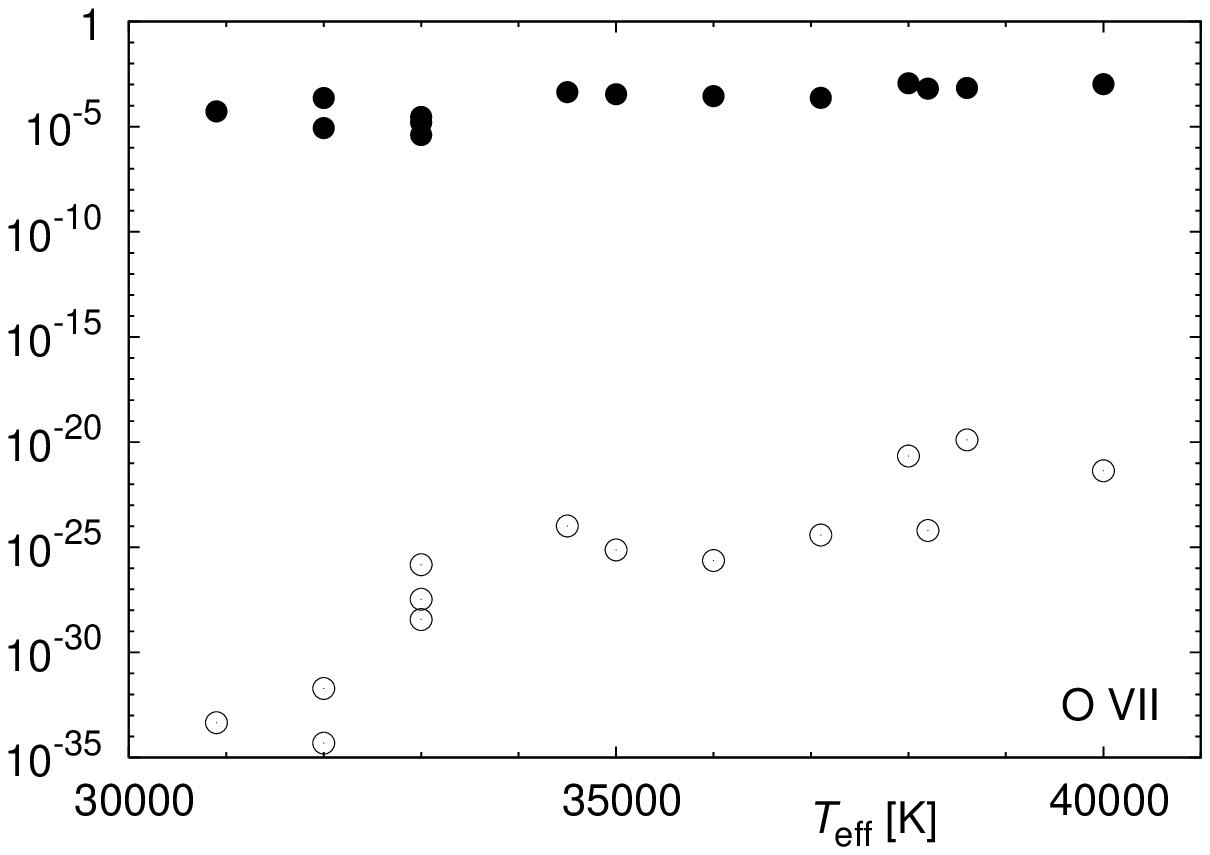}}
\resizebox{0.33\hsize}{!}{\includegraphics{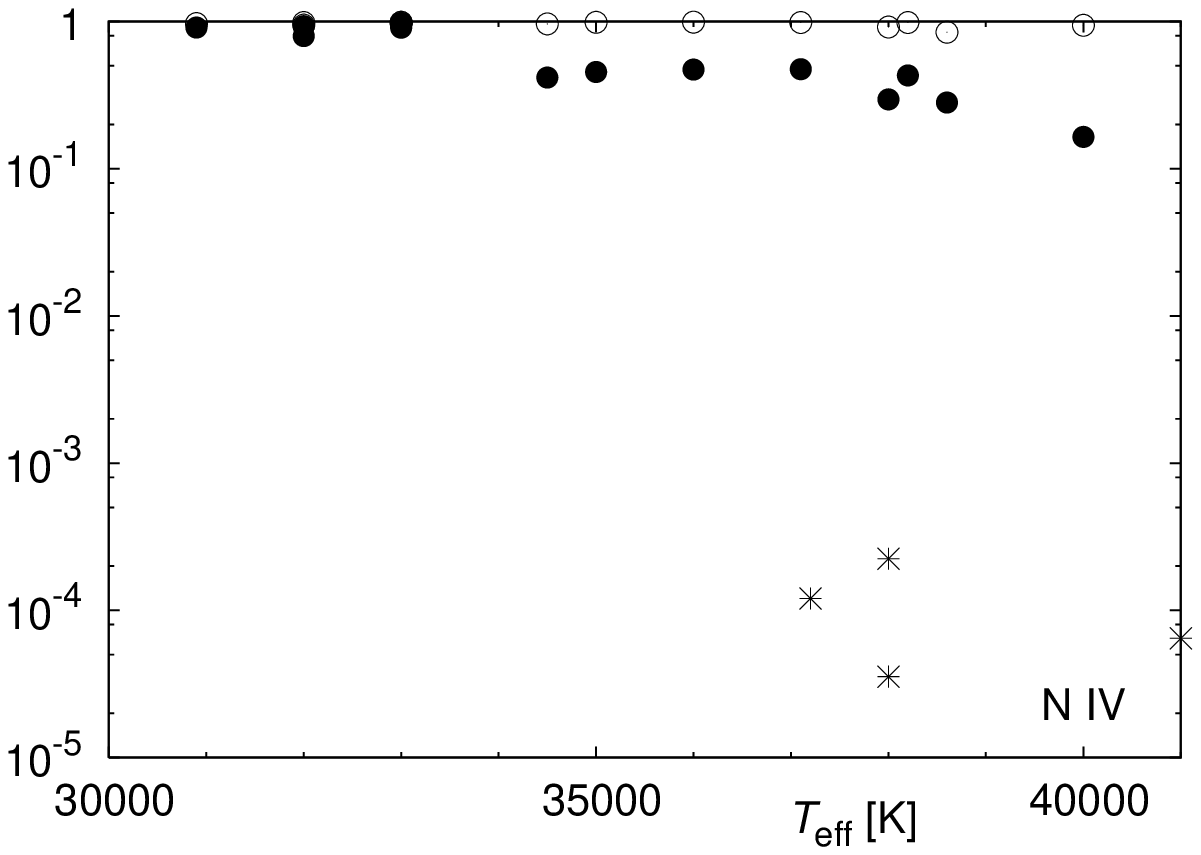}}
\resizebox{0.33\hsize}{!}{\includegraphics{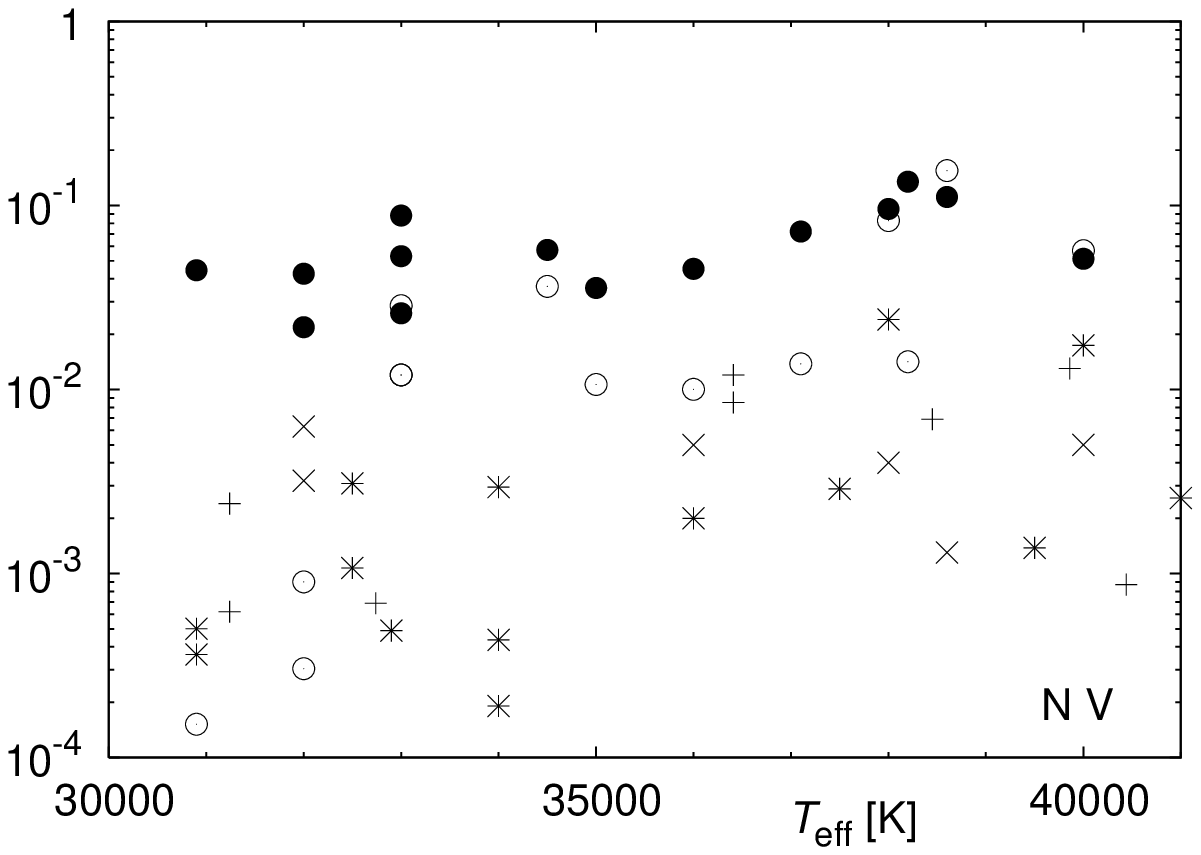}}
\resizebox{0.33\hsize}{!}{\includegraphics{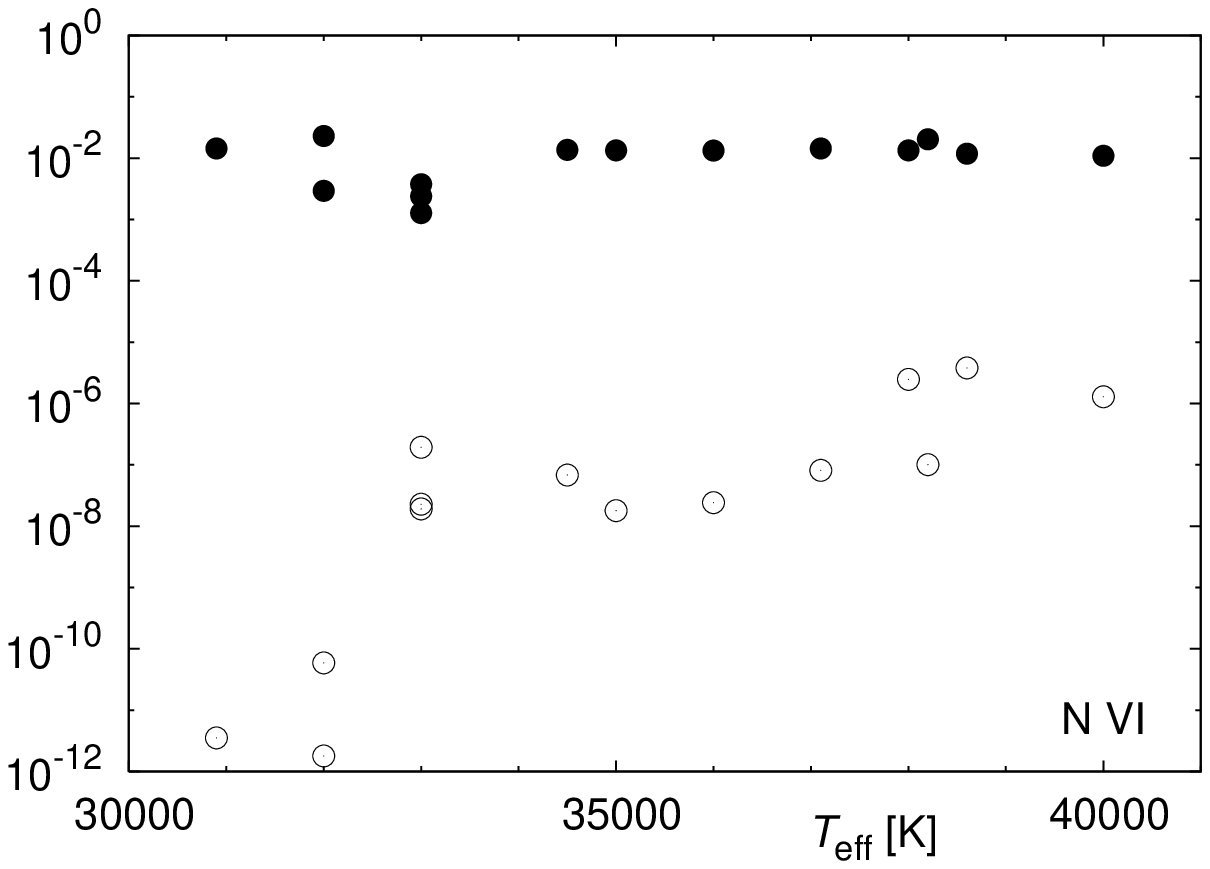}}
\resizebox{0.33\hsize}{!}{\includegraphics{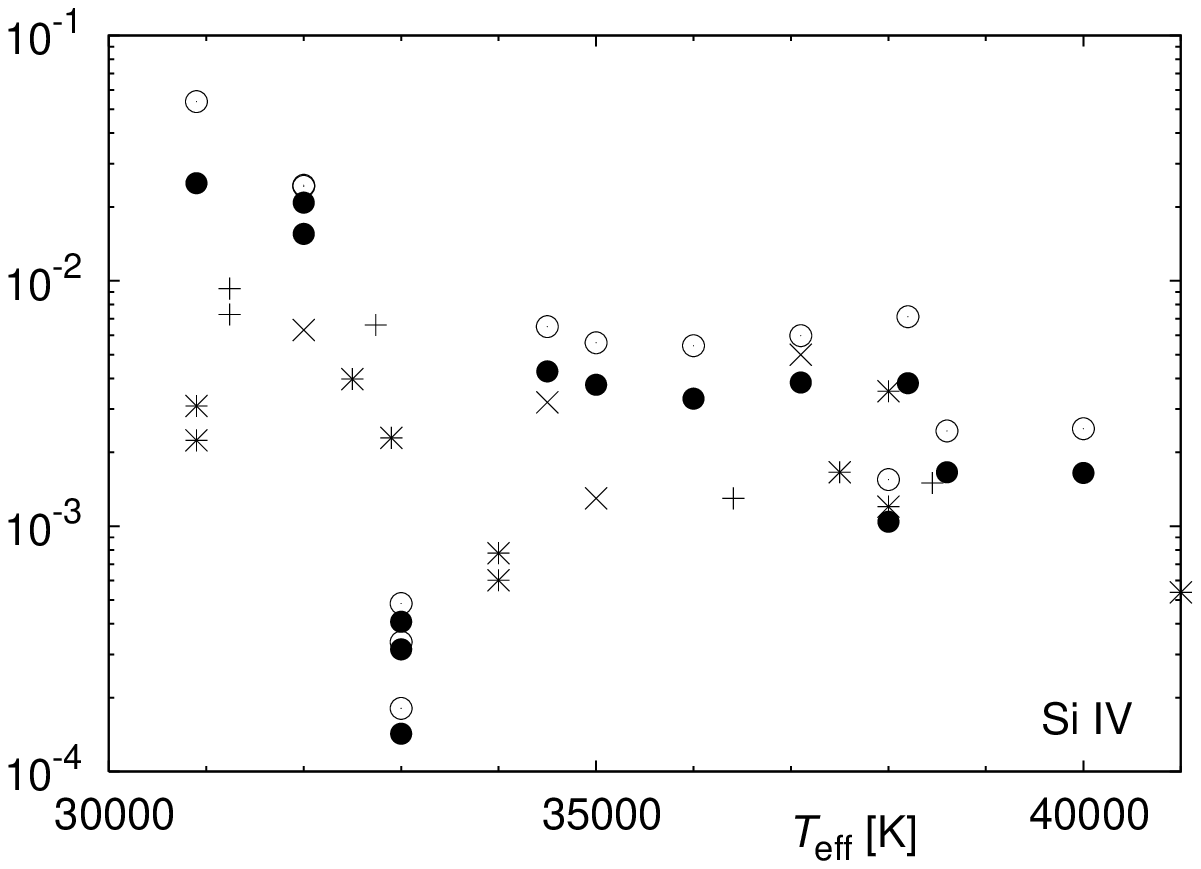}}
\resizebox{0.33\hsize}{!}{\includegraphics{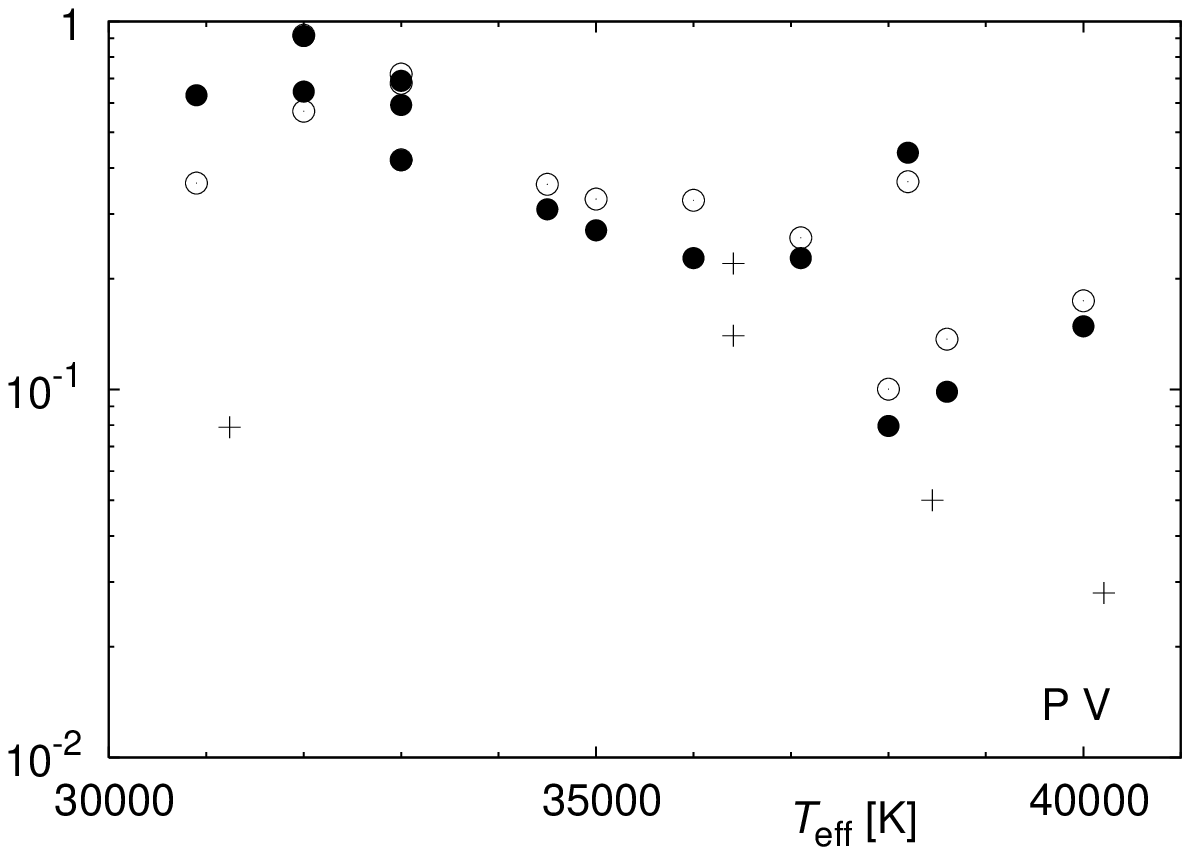}}
\resizebox{0.33\hsize}{!}{\includegraphics{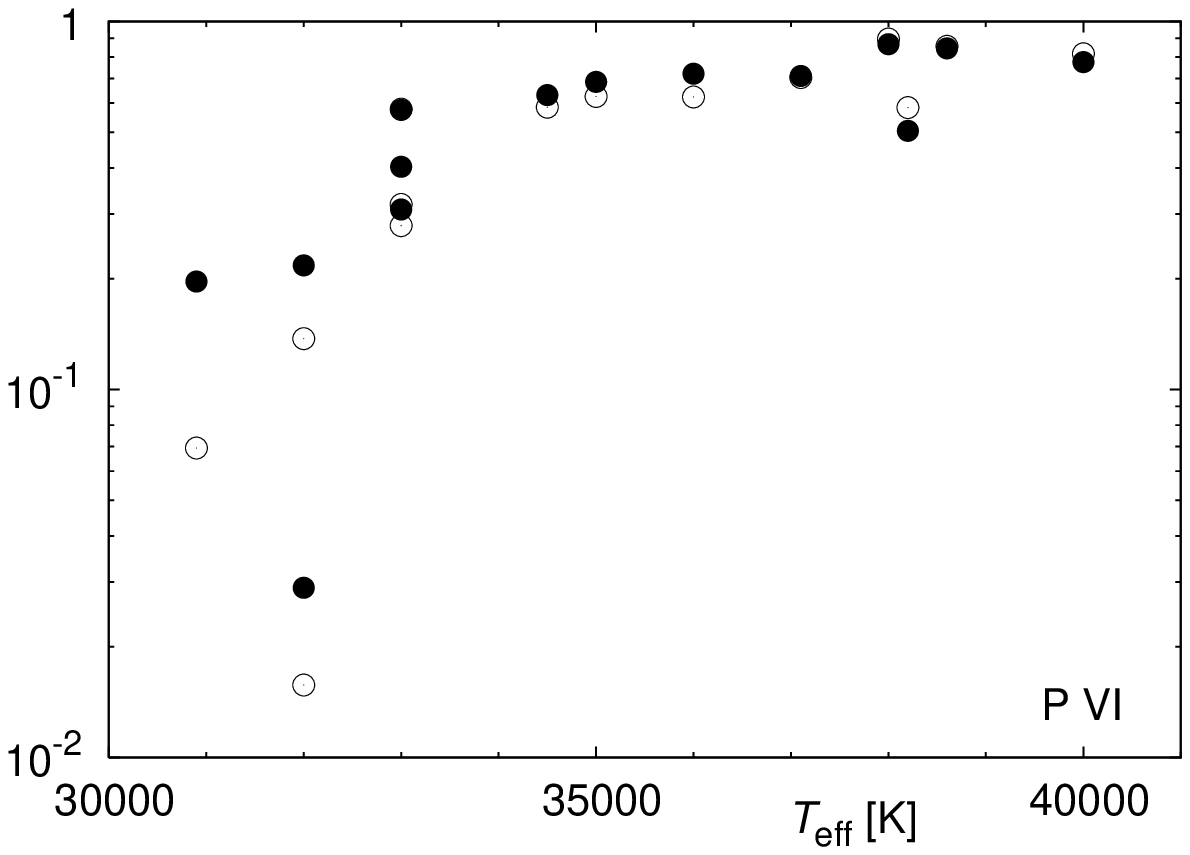}}
\caption{Ionization fractions as a function of the effective temperature for
individual stars from our sample at the point where the radial velocity
$\vr=0.5v_\infty$. Open circles $\odot$ denote values taken from models without
X-ray radiation, filled circles {\lower.3ex\hbox{\LARGE\textbullet}} from models
with $\fx=0.02$. The ionization fraction derived from observations were adopted
from \protect\citet[for LMC stars, plus signs $+$]{maso}, \protect\citet[only
for stars from our sample, crosses $\times$]{hp}, and \citet[asterisks
$\ast$]{lamoc}.}
\label{iontep}
\end{figure*}

To compare the ionization equilibrium for different stars, we plot the
ionization fractions of selected ions for the points where the wind velocity is
equal to $\vr=0.5v_\infty$ (Fig.~\ref{iontep}). The model predictions are
compared with the ionization fractions derived from observations by \cite{hp},
\citet{lamoc} for Galactic stars, and \cite{maso} for stars from the Large
Magellanic Cloud. Stars from the Clouds have generally lower metallicity than
the Galactic ones \citep[e.g.][]{martin}. However, because the dependence of
ionization fractions on metallicity is not strong, we use also these fractions
for our comparison.

\begin{figure}
\centering
\resizebox{0.8\hsize}{!}{\includegraphics{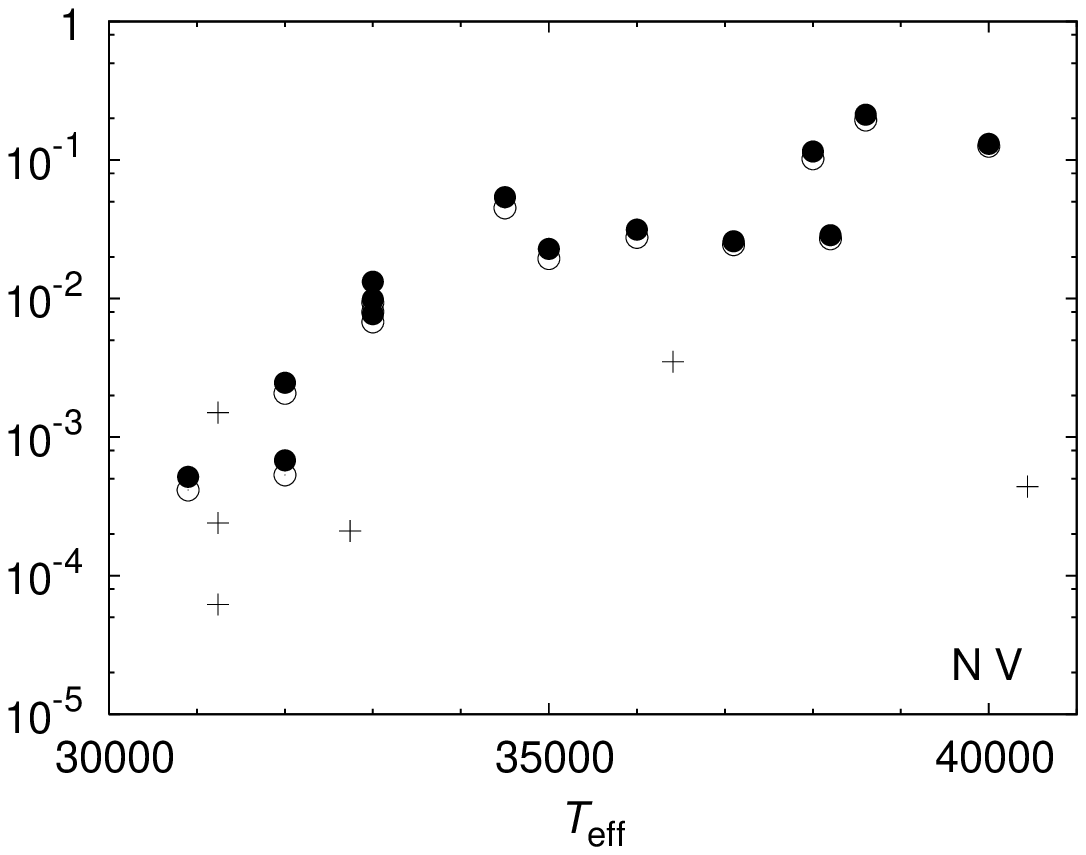}}
\resizebox{0.8\hsize}{!}{\includegraphics{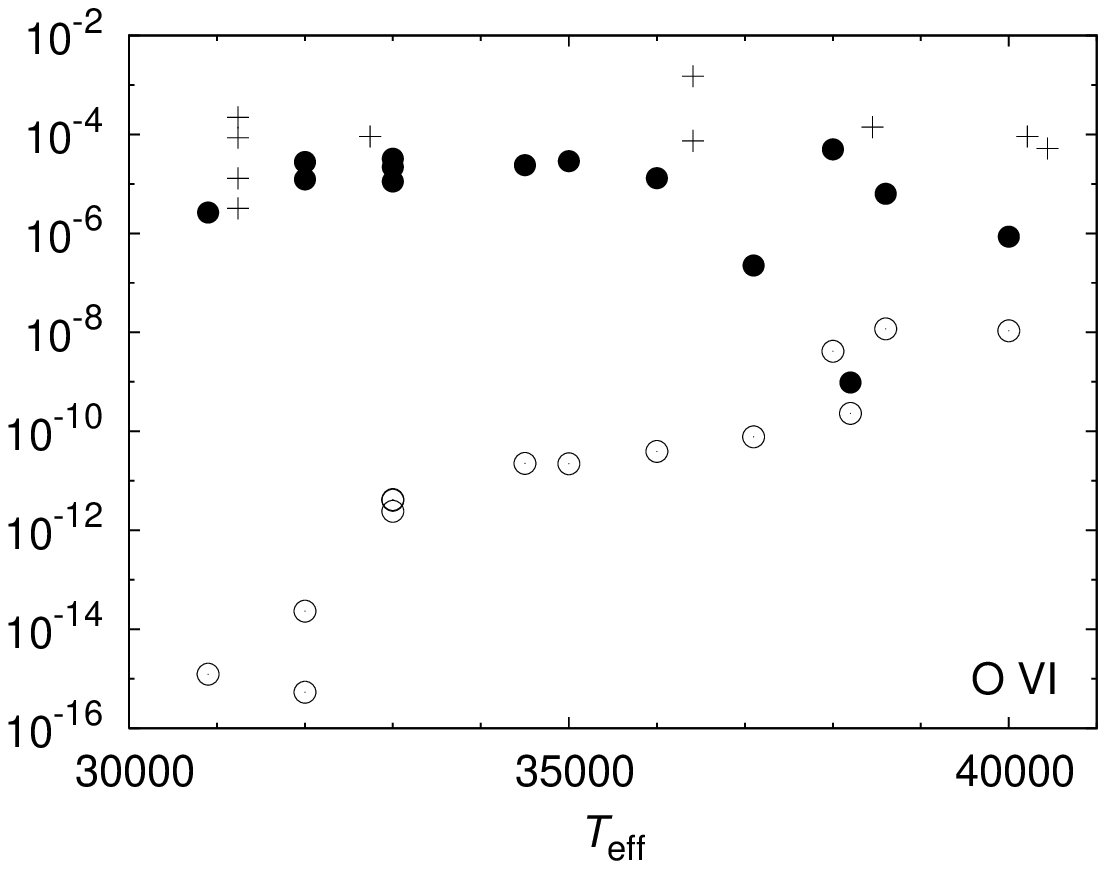}}
\caption{The same as Fig.~\ref{iontep}, however for $\vr=0.05v_\infty$.}
\label{iontepdvac}
\end{figure}

\begin{figure}
\centering
\resizebox{0.8\hsize}{!}{\includegraphics{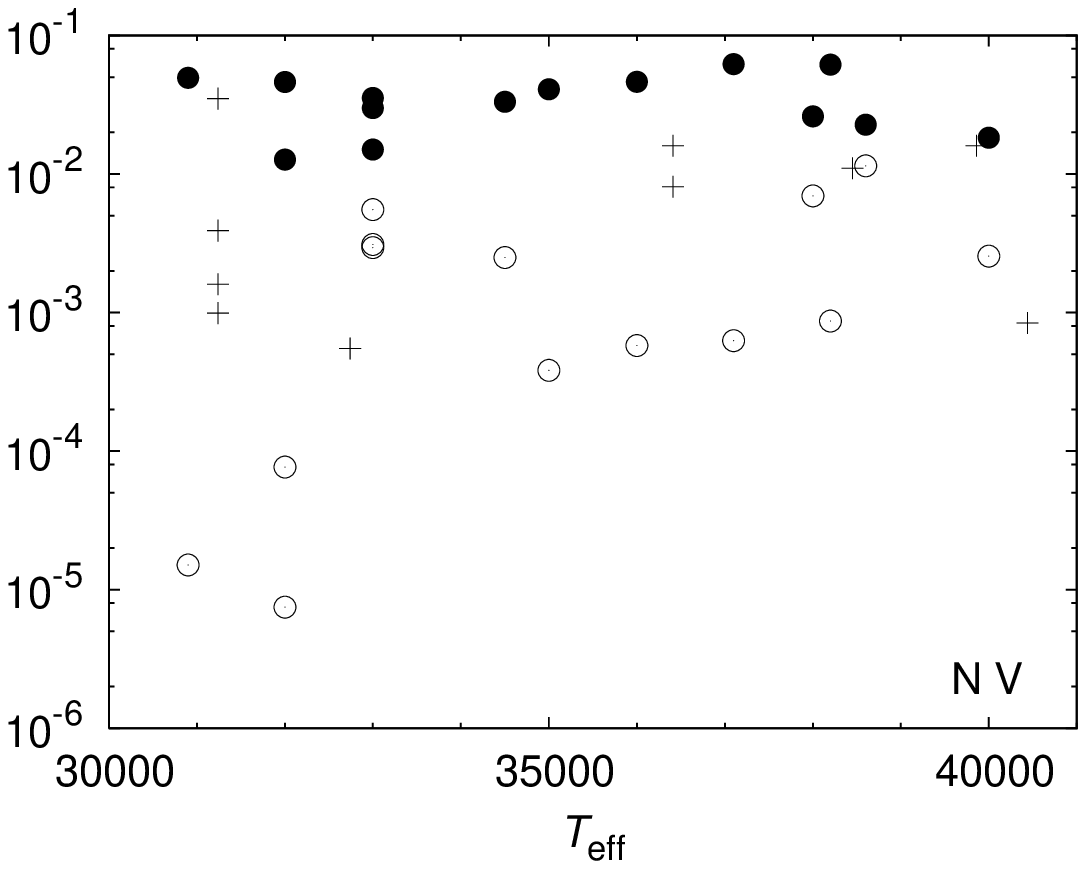}}
\caption{The same as Fig.~\ref{iontep}, however for $\vr=0.8v_\infty$.}
\label{iontepvnek}
\end{figure}

The ionization fractions are influenced not only by the stellar effective
temperature and by local wind parameters, but also by the amount of additional
ionizing X-rays. Consequently, the variations plotted in Fig.~\ref{iontep} are
far from being monotonic. In agreement with MacFarlane et
al.~(\citeyear{macown}), the ionization fractions of the dominant ions are
usually not influenced by X-ray emission. X-rays above all modify the ionization
fractions of minor higher ionization states of a particular atom both due to the
direct and Auger ionization. There is a satisfactory agreement between our
results and models of MacFarlane et al.~(\citeyear{macown}). Note the influence
of X-rays on the ionization balance already very deep in the wind as both
observational and theoretical results show relatively large amount of
\ion{O}{vi} close to the stellar surface for $\vr=0.05\,v_\infty$
(Fig.~\ref{iontepdvac}). However, X-rays do not alter the ionization state of
the dominant and of the lower neighbouring ions there (cf.,
Fig.~\ref{210839ion}).

The influence of X-rays on the ionization structure is stronger for stars with
less dense winds and for stars with weaker flux of the ionizing radiation,
i.e.~for cooler stars.

The presence of \ion{N}{v} lines in the wind spectra is a frequently used
argument for the existence of an additional source of ionization. However, as we
can see from the comparison of the ionization structure derived from
observations and from our models for $v=0.05 v_\infty$, and $v=0.5 v_\infty$
(Figs.~\ref{iontepdvac}, \ref{iontep}), the ionization fraction of \ion{N}{v}
there can be explained just by the detailed NLTE models without a contribution
of X-ray radiation. This was shown already by \citet{pasam}. Only the presence
of \ion{N}{v} in a significant amount at higher wind speeds (see
Fig.~\ref{iontepvnek} and also Fig.~\ref{210839ion}) is caused by the existence
of a strong X-ray source.

\subsection{Phosphorus}

The failure of present NLTE wind models to reproduce the ionization fractions of
\ion{P}{v} derived from observations is one of the main arguments for lower
values of wind mass-loss rates. There is large discrepancy between \ion{P}{v}
ionization fraction derived by us and those derived from observations by
\citet{fuj}, who concluded that either on average the ionization fraction of
\ion{P}{v} is lower than $0.1$ or there is a significant disagreement between
mass-loss rates derived from \ion{P}{v} and H$\alpha$ lines. Some part of this
discrepancy may come from the neglected influence of inhomogeneities on the wind
ionization fractions or radiative transfer \citep{pulamko,mychuch,lidaarchiv}.
Anyway, it is worthwhile to understand whether some part of this discrepancy is
not caused by the fact that \ion{P}{v} is a fragile ion, which is not a dominant
one for any effective temperature, and to test the influence of X-rays on
\ion{P}{v} ionization fraction.

\begin{figure}
\centering
\resizebox{0.8\hsize}{!}{\includegraphics{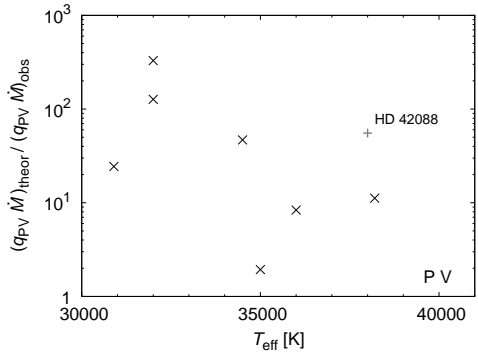}}

\caption{Comparison of the product of the \ion{P}{v} ionization fraction and the
wind mass-loss rate derived from observations \citep{fuj} and predicted ones
(for $\vr=0.5v_\infty$ and $\fx=0.02$). The predicted quantities are much higher
than the observed ones. Gray plus symbol corresponds to the star HD~42088, that
exhibits the ``weak wind problem''.}
\label{p5dmdt}
\end{figure}

Our calculations show that \ion{P}{v} is a dominant ion in the inner part of the
wind only for stars with $T_\text{eff}\lesssim34\,000\,$K (in accordance with
theoretical results discussed by \citealt{fuj}, see right lower panels of
Fig.~\ref{iontep} for the the predicted ionization fraction of \ion{P}{v} and
\ion{P}{iv}). For hotter stars also \ion{P}{vi} becomes an important ionization
stage. However, the presence of shock X-ray radiation does not significantly
modify the ionization fraction of \ion{P}{v}. Consequently, even for hotter
stars for which \ion{P}{v} is not a dominant ion there is a significant
discrepancy between theoretical and observational results, as can be seen from
Fig.~\ref{p5dmdt}. Here we plot the values, which can be in fact derived from
observations, i.e.~the product of the ionization fraction and the mass-loss rate
$q_\text{\ion{P}{v}}\dmdt$, and compare it with the predicted ones. For hotter
stars ($T_\text{eff}\gtrsim35\,000\,$K), the predicted values are on average by
a factor of 7 higher than those derived from observations (with an exception of
the star HD~42088, which displays the ``weak wind problem''). On the other hand,
for cooler stars ($T_\text{eff}\lesssim35\,000\,$K) the predicted values are on
average even about 100 times higher than those derived from observations.

\begin{figure}
\centering
\resizebox{0.8\hsize}{!}{\includegraphics{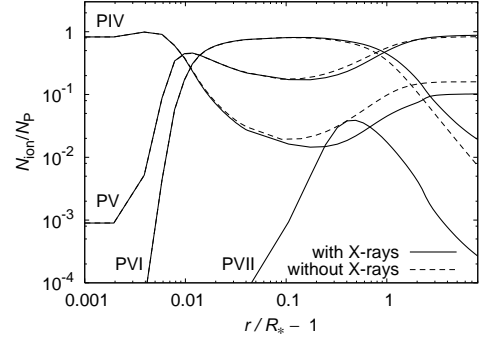}}
\caption{The influence of X-rays on the radial variations of phosphorus
ionization fractions in the wind model of the star HD~203064.
%Kr1:
Note that X-rays do not significantly change \ion{P}{v} ionization fraction
and that \ion{P}{v} dominates for $r\gtrsim2R_*$.}
\label{203064ion}
\end{figure}

The maximum frequency considered in our models (corresponding to 1.7\,keV) is
lower than \ion{P}{v} inner-shell ionization energy threshold 2.2\,keV. To test
our results, we calculated also models with higher maximum frequency considered,
i.e., corresponding to 3.7\,keV (see Fig.~\ref{203064ion}). However, even in
these models the Auger ionization of \ion{P}{v} due to low flux of high energy
X-rays does not significantly modify the ionization fraction of \ion{P}{v}.

\section{Influence of X-rays on hydrodynamic structure}

\begin{figure}
\centering
\resizebox{0.8\hsize}{!}{\includegraphics{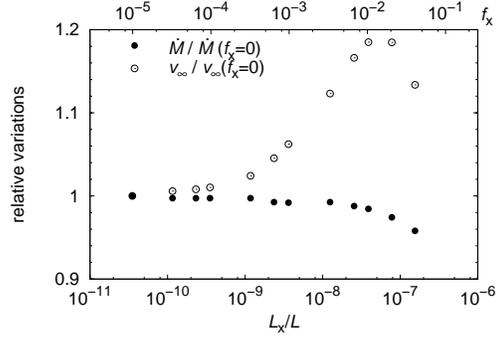}}
\caption{Variations of the mass-loss rate $\dot M$ and the terminal velocity
$v_\infty$ of $\alpha$~Cam with $\fx$ and $L_\text{X}$ ($L_\text{X}$ depends
linearly on $\fx$) relative to their values derived using the models with no
shock X-ray sources.}
\label{lx30614}
\end{figure}

To understand the influence of X-rays on wind parameters, we calculated several
wind models corresponding to $\alpha$~Cam stellar parameters, however with a
different value of $\fx$, i.e. with a different value of X-ray luminosity. The
variations of wind parameters with X-ray luminosity are shown in
Fig.~\ref{lx30614}. As a consequence of assumed dependence of X-ray emissivity
on \fx\ via equation~\eqref{etax}, the emergent X-ray luminosity depends
linearly on $\fx$. For lower X-ray luminosities ($L_\text{X}/L\lesssim10^{-8}$)
both mass-loss rate and terminal velocity are not significantly affected by the
X-ray radiation, since the latter influences only the ionization state of minor
ions (typically VI--VII). This picture changes for higher X-ray luminosities
($L_\text{X}/L\gtrsim10^{-8}$). The mass-loss rate slightly decreases with
increasing X-ray luminosity, whereas the terminal velocity slightly increases.
Changes of wind parameters are caused by modification of ionization equilibrium
due to X-rays. The X-ray radiation shifts the ionization balance to higher
excited states of all elements. Because the number density of minor ionization
state \ion{Fe}{iv}, which is important for the radiative acceleration close to
the star, is lowered, the radiative acceleration decreases. Consequently, the
wind mass-loss rate slightly decreases in the case of $\alpha$~Cam.

Generally, the changes of the wind mass-loss rate due to X-rays are relatively
small. The mass-loss rates for the same star calculated with \fx\ from 0 to 0.04
differ by no more than about 5\%. This is connected with the fact that in our
models the wind mass-loss rate is determined in the region close to the stellar
surface below the critical point. Because this wind region is opaque to X-rays,
the X-rays do not influence the ionization states significantly contributing to
the radiative force there and the mass-loss rates remain basically unaltered.

The change in ionization due to the presence of X-rays does not decrease
the line-force due to the lighter elements. The important lines stay optically
thick and higher excited elements capable to contribute to the radiative force
emerge (e.g., \ion{N}{v}, \ion{O}{v}, \ion{O}{vi}). Thus, the situation in the
outer wind, where the contribution of iron lines is smaller and the influence of
X-rays is higher, is different. The increase of the radiative force (supported
in some cases by the decrease of $\dmdt$) causes the increase of the wind
terminal velocity.

\begin{figure}
\centering
\resizebox{0.8\hsize}{!}{\includegraphics{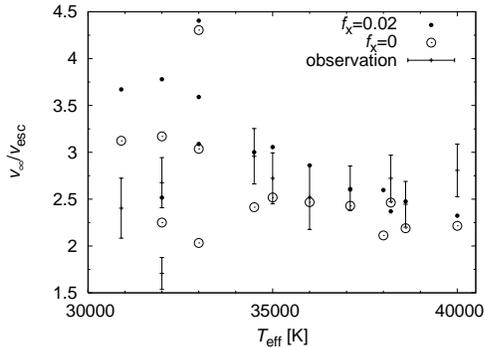}}
\caption{Comparison of the ratio $v_\infty/v_\text{esc}$ calculated with a
theoretical value of $v_\infty$ for $\fx=0.02$,
for neglected X-ray emission ($\fx=0$), and the
ratio $v_\infty/v_\text{esc}$ with $v_\infty$ from observations (observed values
for stars exhibiting weak-wind problem are excluded from this plot).}
\label{vnekvuni}
\end{figure}

For stars with stronger ionizing flux in the far UV region the influence of
X-rays on the wind structure is not so significant. More X-rays are necessary
to modify the ionization fractions of wind driving ions. Consequently, the
increase of the wind terminal velocity is larger for cooler stars
($\Teff\lesssim35\,000\,$K, see Fig.~\ref{vnekvuni}) or for stars with more
tenuous winds.

The energy of artificially included X-rays also transforms to the thermal energy
due to the X-ray absorption. Consequently, wind temperature may be by few
thousands Kelvin higher due to X-rays. However, this effect does not
significantly influence the radiative force.

\section{Consequences for the X-ray line transfer}

Using modern X-ray spectrographs it became possible to resolve the X-ray line
profiles of hot stars. The study of $fir$ lines of He-like ions enabled to
derive radii at which the X-ray emission is generated \citep[e.g.,][]{spravnexp,
milca,raa, leucochce}. It is also possible to predict the shapes of these X-ray
lines 
%Kr1: \citep[e.g.][]{oski}.
\citep[e.g.,][]{neusetrimeradek}.
Most of the predictions assume that the X-ray lines
originate in the hot optically thin environment surrounded by cool wind, which is
optically thick in continuum. These theoretical calculations based on currently
available theoretical mass-loss rates predict asymmetric X-ray line profiles due
to the wind continuum absorption. However, the observed X-ray line profiles are
in most cases symmetric \citep[e.g.,][] {milca,leurez}. This either indicates
that the theoretical wind mass-loss rates are overestimated or this leads to the
discussion on the influence of wind inhomogeneities on the X-ray line transfer
\citep{lidarikala,owosha,oski,ocpor}. Here we study the influence of modified
ionization equilibrium due to hot shock emission on the line transfer in the
X-ray region.

\subsection{Influence of shock emission on the X-ray opacity}

\begin{figure}
\centering
\resizebox{0.8\hsize}{!}{\includegraphics{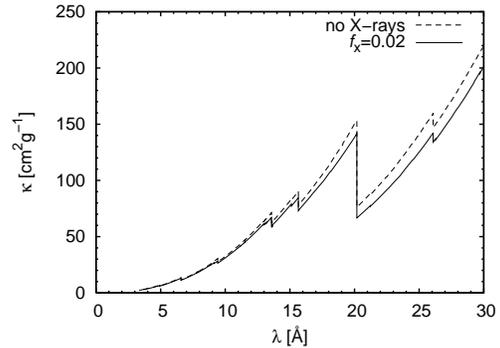}}
\caption{X-ray continuum opacity calculated with and without shock X-ray
emission for $\xi$~Per. Here we give the opacity per unit of mass averaged for
the radii $1.5\,\text{R}_{\odot}<r<5\,\text{R}_{\odot}$. The inclusion of X-rays
leads to a slight decrease of the opacity in the X-ray region.}
\label{xopac}
\end{figure}

The theoretical prediction of the asymmetric shape of the X-ray line profiles is
sensitive to the value of the opacity in the X-ray region. \citet{spravnez}
concluded that a more realistic (lower) value of adopted metallicity leads to
lower continuum opacity in the X-ray region, and consequently to a better
agreement between theoretical and observed X-ray line profiles. Here we test
whether the ionization shift due to the presence of X-rays may affect the X-ray
continuum opacity.

Our calculated opacities are in a good agreement with those presented by
\citet{oski}. Inclusion of shock X-ray emission leads to a slight decrease (on
average by 8\%) of the calculated opacity per unit of mass due to the ionization
shift of ions \ion{He}{ii}, \ion{C}{v}, \ion{N}{iv}, \ion{O}{iv} etc., as shown
in Fig.~\ref{xopac}. This could slightly improve the agreement between predicted
and observed X-ray line shapes, but the decrease is too small to explain the
main part of the discrepancy.

\subsection{Influence of the resonance-line scattering in the cool wind
on the line transfer in the X-ray region}

\begin{figure}
\centering
\resizebox{0.7\hsize}{!}{\includegraphics{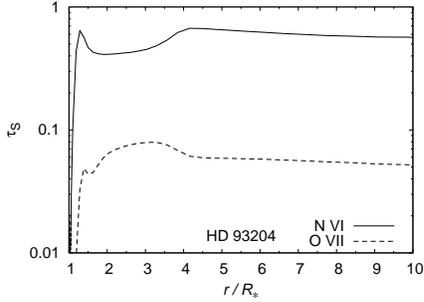}}
\caption{The Sobolev optical depth \eqref{tausobkres} due to the resonance line
scattering due to $1\text{s}^2\,{^1\text{S}}-1\text{s}2\text{p}\,{^1\text{P}}$
lines of helium-like ions \ion{N}{vi} ($\lambda=28.787\,$\AA) and \ion{O}{vii}
($\lambda=21.602\,$\AA) in the cool wind of HD 93204.
The model was calculated for $\fx=0.0094$ that fits the observed X-ray
luminosity.
}
\label{tausobr}
\end{figure}

\citet{tlustocerv} proposed that X-ray emitting regions may be optically thick
in lines. This effect can lead to more symmetric X-ray line profiles, which
better correspond to the observed ones \citep[see also][]{leurez}. However, the
ambient cool wind is usually assumed to absorb the X-ray radiation only in
continuum and the line absorption within the cool wind is being neglected. Here
we test whether this assumption is adequate.

He-like ions \ion{N}{vi} and \ion{O}{vii}, whose lines are observed in the X-ray
spectra of hot stars, are present in a nonneglible amount also in a cool ambient
wind (see Fig.~\ref{iontep}). This is caused by the direct and Auger ionization
of corresponding less ionised ions. The influence of the resonance-line
scattering in the {\em cool} ambient wind on the profiles of X-ray lines emitted
in shocks can be estimated from the optical depth of these lines in the cool
wind. Neglecting the occupation number of upper level $n_i/ g_i \gg n_j/ g_j$,
the Sobolev optical depth in a given line is within the radial streaming
approximation roughly given by \citep[e.g.,][]{cassob}
\begin{equation}
\label{tausobkres}
\tau_\text{S}=\frac{\pi e^2}{m_\mathrm{e}\nu_{ij}}n_if_{ij}
\zav{\frac{\de \vr}{\de r}}^{-1},
\end{equation}
where $\nu_{ij}$ is the line frequency, $f_{ij}$ is the oscillator strength, and
$n_i$, $n_j$ are the number densities of individual states with statistical
weights $g_i$, $g_j$. From this equation it follows that the effect of the
resonance line scattering in the cool wind is effective mainly for stars with
high mass-loss rates or for stars where X-rays more significantly influence the
ionization equilibrium (i.e., stars with large \fx, where higher fractions of
He-like ions emerge).

For stars HD 93204, 162978, 209975, and 210839 the resonance line scattering in
the cool wind is able to influence the emergent X-ray line profiles of
\ion{N}{vi} and in some cases also of \ion{O}{vii} (see Fig.~\ref{tausobr}).
This effect, discussed also by \citet{lidataky}, is sensitive to the properties
of X-ray sources.

\section{Discussion}

\subsection{The dependence of \fx\ on the cooling time}
\label{hustopln}

\begin{figure}
\centering
\resizebox{0.8\hsize}{!}{\includegraphics{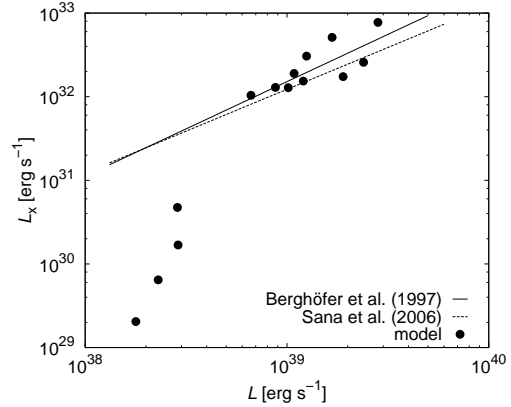}}
\caption{The X-ray luminosity calculated assuming the
dependence of \fx\ on the cooling time via equation~\eqref{promfxrov}
in comparison with observations.}
\label{promfxob}
\end{figure}

%Kr1:
The parameter \fx\ is an important
%Ku1:
free
parameter that determines the wind X-ray
luminosity. As it was used for some models, it is
%Ku1: be
%
possible to
specify \fx\ individually for each star using observed X-ray luminosity.
As with fixed \fx\ we predict steeper $L_\text{X}-L$ relation than observed
(Fig.~\ref{lxlbolhv}), \fx\
%Ku1: that
which
is inversely proportional to the luminosity
would lead to a $L_\text{X}-L$ relation that better corresponds to the observed
one.

%Kr1: The assumption of the fixed \fx\ for all stars may not be adequate.
The inverse proportionality of \fx\ on $L$ (or on the wind density) may be
a more realistic one.
The total
X-ray emissivity is given by the amount of energy dissipated by the shocks and
the fraction of X-ray emitting material (or \fx) may be determined by this
dissipated energy and by the cooling length. In fact, \fx\ may be lower for
stars with dense winds in which the hot material is able to cool down more
efficiently. 
%Kr1: To study this effect,
To correct our results for the dependence of \fx\ on the cooling time,
we calculated wind models where \fx\ is scaled
by the value of the cooling time relatively to the hydrodynamical time-scale
\citep[cf.][]{igor, svetozar}, i.e.~roughly given by
\begin{equation}
\label{promfxrov}
\fx=f_0\epsilon_\text{X}\equiv
f_0\frac{a_\text{H}^2\rho}{(\xi\rho)^2\int_0^\infty
\Lambda(T_\text{X})\,\text{d}\nu}\frac{\vr}{r},
\end{equation}
where we assume $f_0=0.7$. Since according to \eqref{promfxrov} $\fx\sim r$,
$L_\text{X}$ now increases ad infinitum. To avoid a possible divergence of
$L_\text{X}$ we limit \fx\ in \eqref{promfxrov} by a certain value, namely
$\fx\leq0.1$ as in the outer regions shocks may become adiabatic or weaker
\citep{felkupa,felpulpal}. The derived relation between $L_\text{X}$ and $L$ in
Fig.~\ref{promfxob} shows a good agreement with observations indicating that the
variable cooling time may indeed be an important property of wind shocks.

\subsection{The weak wind problem}
\label{slabo}

\begin{figure}
\centering
\resizebox{0.8\hsize}{!}{\includegraphics{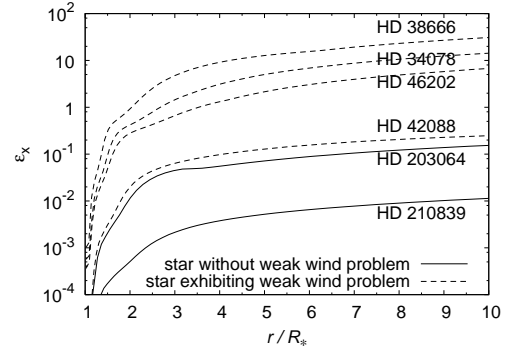}}
\caption{The radial variations of $\epsilon_\text{X}$ (see equation
\eqref{promfxrov}) for selected stars.}
\label{ochladel}
\end{figure}

Comparison of mass-loss rates derived from observation with the predicted ones
(Table~\ref{obvitpar}) shows that there is a good agreement between observation
and theory for stars with large mass-loss rates $\dmdt\gtrsim10^{-7}\,\msr$
(provided that the observations are not significantly influenced by clumping).
The situation is markedly different for stars with lower mass-loss rates
$\dmdt\lesssim10^{-7}\,\msr$ (except HD~149757). The predicted mass-loss rates
are generally by order of magnitude higher than those derived from observations.
These stars exhibit the so-called ``weak wind problem", whose solution is not
yet known (\citealt{bourak,martin}, M05, \citealt{nlteii}).

The X-ray activity of these stars is higher than that of stars which do not
exhibit the weak wind problem. Within our models this means that their X-ray
luminosity predicted assuming a fixed filling factor is by two to three orders
of magnitude lower than the observed one (see Fig.~\ref{lxlbolhv}). However, the
X-ray emission of these stars may still originate in the wind, because their
X-ray luminosity is lower than the wind kinetic energy lost per unit of time
$1/2\dot Mv_\infty^2$ \citepalias{martclump}. Note that due to low wind density
of these stars the cooling length may become comparable with the hydrodynamical
scale (as indicates the plot of $\epsilon_\text{X}$ in Fig.~\ref{ochladel}).
Consequently, there is a possibility that once the wind of these stars is heated
by the shocks it is not able to cool down sufficiently, and remains hot and
unaffected by the radiative acceleration. This can be a reasonable explanation of
the ``weak wind problem" and high X-ray luminosity $L_\text{X}$ of these stars.
A similar scenario was invoked by \citet{cobecru} to explain the X-ray spectrum
of $\beta$ Cru.

\subsection{The metallicity dependence}
\label{zkap}

To discuss the influence of metallicity on $L_\text{X}$ we recalculated our
models with lower value of the mass fraction of heavier elements $Z$. Because
the dependence of the free parameter \fx\ on metallicity is not known, we
calculated these models with fixed \fx.

\begin{figure}
\centering
\resizebox{0.7\hsize}{!}{\includegraphics{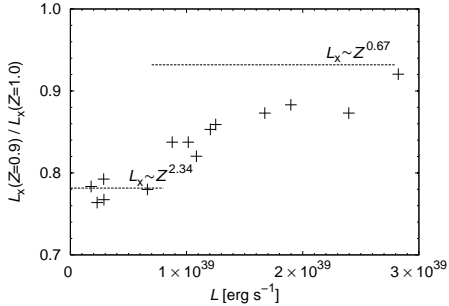}}
\caption{The ratio between the calculated X-ray fluxes from models with
metallicities $Z=0.9$ and $Z=1$ (for the same $\fx=0.02$).
The dashed lines denote the approximate relations
(equations \eqref{ztenky}, \eqref{ztlusty}).}
\label{lxzlx}
\end{figure}

The derived variations of $L_\text{X}$ with $Z$
given in Fig.~\ref{lxzlx}
can be easily interpreted. The mass-loss rate depends on
metallicity roughly as $\dot M\sim Z^{0.67}$ \citep{nlteii}. The X-ray emission
originates mainly from lines of heavier elements, thus we can assume that the
emissivity per 
unit of mass
depends on the metallicity roughly as
$\Lambda_\nu(T_\text{X})\sim Z$. Consequently, from the relations of
\citetalias{oskal} (see Section \ref{kaplxskal}) the optically thin X-ray
luminosity scales with the metallicity as
\begin{equation}
\label{ztenky}
L_\text{X}\sim Z^{2.34}.
\end{equation}

In the optically thick case we can assume that for higher frequencies
($\nu\gtrsim1\times10^{17}\,$Hz) the X-ray opacity $\kappa_\text{X}$ is mostly
due to direct and inner-shell ionization of heavier elements. As a result of
this $\kappa_\text{X}\sim Z$, and in the optically thick case
the relations of \citetalias{oskal} yield $L_\text{X}\sim\dot
M\Lambda_\nu /\kappa_\text{X}$, i.e.
\begin{equation}
\label{ztlusty}
L_\text{X}\sim Z^{0.67}.
\end{equation}

The relations \eqref{ztenky} and \eqref{ztlusty} are derived assuming that
\fx\ does not depend on the metallicity. However, lower value of
$\Lambda$ for lower metallicities could effectively mean that the cooling time
of the gas in the post-shock region is longer. Consequently, \fx\ may become
higher for lower metallicities. In such a case we can expect the dependence of
\fx\ on the metallicity roughly as $\fx\sim Z^{-1}$. Within this approximation
the total X-ray luminosity would scale as
\begin{subequations}
\begin{align}
L_\text{X}&\sim Z^{1.3} & \text{(optically thin case),}\\
L_\text{X}&\sim Z^{-0.3} & \text{(optically thick case)}.
\end{align}
\end{subequations}

\subsection{Multicomponent effects}

The multicomponent effects due to inefficient collisions between individual wind
particle components may significantly influence structure of low-density winds
\citep[e.g.,][] {treni,nlteii}. The Coulomb frictional force between wind
components depends on the charge of these components, which may be influenced by
the presence of X-rays. However, our tests showed that such influence is in most
cases insignificant, because the X-rays typically modify the fraction of minor
ionization stages only.

We tested whether the multicomponent effects are important for stars exhibiting
the "weak-wind problem". For this purpose we treated heavier elements as
individual components \citep [as in][] {nlteii}. We concluded that phosphorus
due to its low abundance may decouple from the wind of HD~34078 and HD~38666,
and the velocity difference between phosphorus and hydrogen is high also in the
wind of HD~46202. For these stars the multicomponent effects and the ionization
shift induced by it may be important.

\subsection{Model simplifications}

The inclusion of additional X-ray sources into our NLTE wind models was done
assuming a simplified shock picture, in reality these sources may have slightly
different properties. For example, the X-ray source can not be described just by
one temperature \citep{felkupa}. However, the main result of this paper, i.e.,
that the presence of X-rays does not significantly influence the wind parameters
of O stars was justified for a broad range of shock parameters. On the other
hand, the predicted ionization structure is much more sensitive to properties of
X-ray sources.

There is a growing observational evidence that hot stars winds are clumped
\citep{bourak,martin,pulchuch}. We postpone the effect of clumping on the wind
structure for a special study. The clumping may influence not only the
comparison of ionization fractions derived from observation and from the
theoretical models, but also modify the radiative force and the mass-loss rates
\citep{mychuch,lijana}. On the other hand, the effect of porosity
\citep{owosha,oski} may lead to decrease of the effective opacity in the X-ray
region. This may affect the total X-ray luminosity, but as the porosity affects
also the predicted wind parameters \citep{mychuch} and as its existence in the
wind is a matter of discussion nowadays, we do not include it into our models.
Note that the present approach of the inclusion of both clumping and porosity
into wind models is based on {\em free} parameters whose values are not very
well observationally nor theoretically constrained.

\section{Conclusions}

We studied an effect of additional X-ray emission on our NLTE wind models. To
this end, we assumed that some part of the wind material is heated to a very
high temperature (of order of $10^6\,$K) and emits X-rays, and we studied their
influence on wind structure.

Using our wind models we derived the same scaling of the total X-ray luminosity
$L_\mathrm{X}$ with the mass-loss rate $\dot M$ as \citet {oskal}. For stars
with optically thick winds most of X-rays are absorbed in the wind, and the
X-ray luminosity scales with wind mass-loss rate roughly as
$L_\text{X}\sim\dmdt$. For stars with optically thin wind in the X-ray region
the X-ray luminosity scales as $L_\text{X}\sim\dmdt^2$.

The models with fixed filling factor \fx\ can roughly explain the observed
$L_\text{X}/L$ relation of stars with optically thick wind in the X-ray region.
The remaining deviation between the observed slope of the $L_\text{X}/L$
relation and that derived using fixed \fx\ can be the result of the dependence
of the shock cooling time on wind density. The stars with optically thin wind
exhibiting the "weak wind problem" emit more X-rays than predicted by a simple
optically thin scaling. Their enhanced X-ray emission and the "weak wind
problem" itself may be caused by large wind shock cooling time, comparable with
the hydrodynamical time scale.

The X-rays influence the ionization state in the whole wind. However, a
significant amount of X-rays is emitted only in highly supersonic wind regions.
These X-rays mostly do not penetrate close to the stellar surface, consequently
the ionization changes induced by X-rays there are not large enough to change
the radiative force there significantly. Because the mass-loss rate is
determined in the region close to the stellar surface, the presence of X-rays
does not significantly change its value. On the other hand, the terminal
velocity is determined in the outer wind regions where the influence of X-rays
is more important. Consequently, the wind terminal velocity may be slightly
affected by X-rays, especially for stars with weaker ionizing continua (with
$T_\text{eff}\lesssim35\,000\,$K).

Although the Auger processes are important for the ionization balance in the
outer wind regions, the non-negligible presence of higher ionization stages
(sometimes also called "superions") is a consequence of also a direct
ionization.

We compared the model ionization fractions with those derived from observations.
For higher ionization stages the inclusion of X-rays leads to better agreement
between theory and observations. However, for many ions a significant
discrepancy between theory and observations still remains. The discrepancy
between the mass-loss rates derived from the H$\alpha$ line or radio emission
and \ion{P}{v} line profiles led \citet{fuj} to conclusion that real wind
mass-loss rates of hot stars are much lower than the ``classical" ones. This
conclusion is sensitive to the assumed \ion{P}{v} ionisation fraction. We have
shown that for stars with $T_\text{eff}\gtrsim34\,000\,$K the \ion{P}{vi} ion
becomes a dominant one, which gives a better agreement for hotter stars. But
even for these hotter stars a significant disagreement between theory and
observations 
%Kr1: still
remains.
%Kr1:
We
%Ku1: do not found
did not find
any significant influence of X-rays on \ion{P}{v}
ionization fraction. Our calculations suggest that the effect of X-rays cannot
be the reason the \ion{P}{v} mass-loss rates are low. This problem is especially
severe for stars with $T_\text{eff}\lesssim34\,000\,$K for which we
predict \ion{P}{v} to be the dominant ionization stage in the wind.

We studied the influence of modified ionization equilibrium due to shock
emission on the X-ray line transfer. While continuum absorption by the cool wind
in the X-ray domain is not significantly affected by X-rays itself, additional
line opacity sources in a cool wind emerge \citep[see also][]{lidataky}. This
may affect the line
%Kr1: profile shapes
profiles of more abundant elements (C, N, and O)
observed in the X-ray spectra.

We can conclude that although the inclusion of X-rays into NLTE wind models
leads to a better agreement between theory and observation, it can not explain
the most painful problems of present hot star wind research, i.e., lower
mass-loss rates derived from observation with taking into account clumping, and
too-low ionization fractions of \ion{P}{v} derived from observations.

\section*{Acknowledgements}
We thank to the anonymous referee for his/her comments that helped to improve
the manuscript. We thank to Dr.~J.~Puls for the discussion and providing us
atomic data, and to Prof. S.~Owocki, Drs.~L.~Oskinova and V.~Votruba for the
discussion. This research has made use of NASA's Astrophysics Data System and
the SIMBAD database, operated at CDS, Strasbourg, France. This work was
supported by grant GA AV \v{C}R B301630501, and later also by a grant GA \v{C}R
205/08/0003.

\appendix
\section{Calculation of diagonal elements of \boldmath$\Psi_\nu$}
\label{usti}

The equation of the radiative transfer \eqref{bezpravi} is solved using the LU
decomposition,
\begin{equation}
\Psi^{-1}_\nu = {\mathsf L}{\mathsf U},
\end{equation}
where $\mathsf L$ is a product of permutation and lower triangular matrices with
1 subdiagonal, and $\mathsf U$ is upper triangular matrix with 2 superdiagonals.
In such a case the inversion matrix $\Psi_\nu$ can be easily calculated (Ralston
\citeyear{ral}) as
\begin{equation}
\Psi_\nu ={\mathsf U}^{-1}{\mathsf L}^{-1},
\end{equation}
where ${\mathsf U}^{-1}$ and ${\mathsf L}^{-1}$ are upper and lower diagonal
matrices. If we denote ${\mathsf u}\equiv{\mathsf U}^{-1}$ and 
\begin{align}
{\mathsf u}_{i,i}=&{\mathsf U}_{i,i}^{-1},\\
{\mathsf u}_{i,j}=&
   -\frac{{\mathsf u}_{i,j-1}{\mathsf U}_{j-1,j}+
          {\mathsf u}_{i,j-2}{\mathsf U}_{j-2,j}}{{\mathsf U}_{j,j}},
\end{align}
and similarly for ${\mathsf L}^{-1}$. From this, diagonal elements of
$\Psi_\nu$
can be easily calculated.

Instead of using matrix inversion it is possible to derivate \eqref{bezpravi}
with respect to the occupation number $N_i$ and calculate partial derivatives
${\partial J_\nu}/{\partial N_j}$ as a solution of derived system of linear
equations. Although this would be likely fast solution (remember that since we
employ LU decomposition for the solution of equation \ref{bezpravi}, we can use
this solution procedure to solve other equations with the same matrix), we would
have to solve these equations for each $N_j$, and this would be very time
consuming.

%%%%%%%%%%%%%%%%%%%%%%%%%%%%%%%%%%%%%%%%%%%%%%%%%%%%%%%%%%%%%%%%%%%%%%%%
\newcommand{\NRT}[1]{in Numerical Radiative Transfer, W. Kalkofen ed.,
        Cambridge Univ. Press, p.~#1}

\end{document}